\documentclass[sigconf]{acmart}

\usepackage{amsmath,amsfonts} %amssymb
\usepackage[ruled,vlined,noend]{algorithm2e}
\usepackage{graphicx}
\usepackage{textcomp}
\usepackage{xcolor}
\usepackage{multicol}
\usepackage{multirow}
\usepackage{url}
\usepackage{listings}
\usepackage[utf8]{inputenc}
\usepackage{bm}
\usepackage{subcaption}
\usepackage{gensymb}
\usepackage{tikz}
\usepackage{soul}
\usepackage{ragged2e}
\usepackage{enumitem}
\usepackage[para]{footmisc}

\DeclareRobustCommand\shortredline
{\tikz[baseline=0.0ex]\draw[red, line width=0.8pt] (0,0.1) -- (0.2,0.1);}
\DeclareRobustCommand\lightgreybar
{\tikz[baseline=0.2ex]\fill[lightgrey] (0,0) rectangle (0.24,0.24);}
\DeclareRobustCommand\blackbar
{\tikz[baseline=0.2ex]\fill[black] (0,0) rectangle (0.24,0.24);}

\usetikzlibrary{decorations.pathreplacing,calc}
\newcommand{\tikzmark}[1]{\tikz[overlay,remember picture] \node (#1) {};}
\newcommand*{\AddNote}[4]{%
    \begin{tikzpicture}[overlay, remember picture]
        \draw [decoration={brace,amplitude=0.5em},decorate,ultra thick,forestgreen]
            ($(#3)!(#1.north)!($(#3)-(0,1)$)$) --  
            ($(#3)!(#2.south)!($(#3)-(0,1)$)$)
                node [align=center, text width=1.5cm, pos=0.5, anchor=west] {#4};
    \end{tikzpicture}
}%

\newcommand{\STAB}[1]{\begin{tabular}{@{}c@{}}#1\end{tabular}}

\newcommand{\ie}{i.e.,}
\newcommand{\eg}{e.g.,}
\newcommand{\wrt}{w.r.t.}

\newcommand{\oursystem}{PPFL\xspace}
\newcommand{\oursystemlong}{Privacy-preserving Federated Learning\xspace}
\newcommand{\dnn}{DNN\xspace}
\newcommand{\dnns}{DNNs\xspace}
\newcommand{\fl}{FL\xspace}

\newcommand{\mysubsubsection}[1]{\vspace{4pt} \noindent \textbf{#1}}

\definecolor{darkgreen}{RGB}{0,100,0}
\definecolor{forestgreen}{RGB}{34,139,34}
\definecolor{lightgrey}{RGB}{204, 204, 204} % Hex #999999
\definecolor{darkgrey}{RGB}{51, 51, 51} % Hex #333333
\definecolor{black}{RGB}{0, 0, 0} % Hex #000000
\definecolor{red}{RGB}{204,0,51} % Hex #CC0033

\author{Fan Mo}
\affiliation{Imperial College London}
\authornote{Work performed while at Telef\'onica Research.}
\author{Hamed Haddadi}
\affiliation{Imperial College London}
\author{Kleomenis Katevas}
\affiliation{Telef\'onica Research}
\author{Eduard Marin}
\affiliation{Telef\'onica Research}
\author{Diego Perino}
\affiliation{Telef\'onica Research}
\author{Nicolas Kourtellis}
\affiliation{Telef\'onica Research}

\ccsdesc[500]{Security and privacy~Privacy protections}
\ccsdesc[300]{Computing methodologies~Distributed algorithms}
\ccsdesc[300]{Security and privacy~Distributed systems security}

\hypersetup{colorlinks,linkcolor={black},citecolor={black},urlcolor={black}}

\begin{document}

% copyright
\copyrightyear{2021} 
\acmYear{2021} 
\setcopyright{acmlicensed}\acmConference[MobiSys '21]{The 19th Annual International Conference on Mobile Systems, Applications, and Services}{June 24-July 2, 2021}{Virtual, WI, USA}
\acmBooktitle{The 19th Annual International Conference on Mobile Systems, Applications, and Services (MobiSys '21), June 24-July 2, 2021, Virtual, WI, USA}
\acmPrice{15.00}
\acmDOI{10.1145/3458864.3466628}
\acmISBN{978-1-4503-8443-8/21/07}

\title{PPFL: Privacy-preserving Federated Learning with Trusted Execution Environments}

\begin{abstract}
\label{sec:abstract}

We propose and implement a Privacy-preserving Federated Learning ($PPFL$) framework for mobile systems to limit privacy leakages in federated learning. Leveraging the widespread presence of Trusted Execution Environments (TEEs) in high-end and mobile devices, we utilize TEEs on clients for local training, and on servers for secure aggregation, so that model/gradient updates are hidden from adversaries.
Challenged by the limited memory size of current TEEs, we leverage greedy layer-wise training to train each model's layer inside the trusted area until its convergence. 
The performance evaluation of our implementation shows that $PPFL$ can significantly improve privacy while incurring small system overheads at the client-side.
In particular, $PPFL$ can successfully defend the trained model against data reconstruction, property inference, and membership inference attacks.
Furthermore, it can achieve \emph{comparable} model utility with fewer communication rounds (0.54$\times$) and a similar amount of network traffic (1.002$\times$) compared to the standard federated learning of a complete model.
This is achieved while only introducing up to $\sim$15\% CPU time, $\sim$18\% memory usage, and $\sim$21\% energy consumption overhead in $PPFL$'s client-side.

\end{abstract}
\maketitle

%%%%%%%%%%%%%%%%%%%%%
% main texts
\section{Introduction}
\label{sec:introduction}

Training deep neural networks (\dnns) on multiple devices locally and building an aggregated global model on a server, namely federated learning (\fl), has drawn significant attention from academia (\eg~\cite{geyer2017differentially, kairouz2019advances, liu2020secure}) and industry, and is even being deployed in real systems (\eg~Google Keyboard~\cite{bonawitz2019towards}).
Unlike traditional machine learning (ML), where a server collects all user data at a central point and trains a global model, in \fl, users only send the locally updated model parameters to the server.
This allows training a model without the need for users to reveal their data, thus preserving their privacy.
Unfortunately, recent works have shown that adversaries can execute attacks to retrieve sensitive information from the model parameters themselves~\cite{zhu2019deep, melis2019exploiting, geiping2020inverting, hitaj2017deep}.
Prominent examples of such attacks are data reconstruction~\cite{hitaj2017deep, geiping2020inverting} and various types of inference attacks~\cite{hitaj2017deep, melis2019exploiting}.
The fundamental reason why these attacks are possible is because as a \dnn learns to achieve their main task, it also learns irrelevant information from users' training data that is inadvertently embedded in the model~\cite{yeom2018privacy}. 
Note that in \fl scenarios, such attacks can be launched both at server and client sides.

Motivated by these attacks, researchers have recently introduced several countermeasures to prevent them.
Existing solutions can be grouped into three main categories depending on whether they rely on:
(i) homomorphic encryption (\eg~\cite{aono2017privacy, liu2020secure}),
(ii) multi-party computation (\eg~\cite{bonawitz2017practical}), or
(iii) differential privacy (\eg~\cite{geyer2017differentially, mcmahan2018learning, dwork2014algorithmic}). While homomorphic encryption is practical in both high-end and mobile devices, it only supports a limited number of arithmetic operations in the encrypted domain.
Alternatively, the use of fully homomorphic encryption has been employed to allow arbitrary operations in the encrypted domain, thus supports ML.
Yet, this comes with too much computational overhead, making it impractical for mobile devices~\cite{naehrig2011can, sealcrypto}. Similarly, multi-party computation-based solutions incur significant computational overhead.
Also, in some cases, differential privacy can fail to provide sufficient privacy as shown in~\cite{melis2019exploiting}.
Furthermore, it can negatively impact the utility and fairness of the model~\cite{jayaraman2019relaxations, bagdasaryan2019differential}, as well as the system performance~\cite{testuggine2020opacus, subramani2020enabling}.
Overall, none of the existing solutions meets all requirements, hampering their adoption.

More recently, the use of hardware-based Trusted Execution Environments (TEEs) has been proposed as a promising way to preclude attacks against \dnn model parameters and gradients.
TEEs allow to securely store data and execute arbitrary code on an untrusted device almost at native speed through secure memory compartments.
All these advantages -- together with the recent commoditization of TEEs both in high-end and mobile devices -- make TEEs a suitable candidate to allow fully privacy-preserving ML modeling.
However, in order to keep the Trusted Computing Base (TCB) as small as possible, current TEEs have limited memory. This makes it impossible to simultaneously place all DNN layers inside the TEE.
As a result, prior work has opted for using TEEs to conceal only the most sensitive DNN layers from adversaries, leaving other layers unprotected~\cite{gu2018securing, mo2020darknetz}.
While this approach was sufficient to mitigate some attacks against traditional ML where clients obtain only the final model, in \fl scenarios the attack surface is significantly larger. \fl client devices are able to observe distinct snapshots of the model throughout the training, allowing them to realize attacks at different stages~\cite{melis2019exploiting, hitaj2017deep}.
Therefore, it is of utmost importance to protect all \dnn layers using the TEE.

In this paper, we propose \oursystemlong (\oursystem), the first practical framework to \emph{fully} prevent private information leakage at both \emph{server} and \emph{client-side} under \fl scenarios.
\oursystem is based on \emph{greedy layer-wise} training and aggregation, overcoming the constraints posed by the limited TEE memory, and providing comparable accuracy of complete model training at the price of a tolerable delay.
Our layer-wise approach supports sophisticated settings such as training one or more layers (block) each time, which can potentially better deal with heterogeneous data at the client-side and speed up the training process.

To show its feasibility, we implemented and evaluated a full prototype of \oursystem system including server-side (with Intel SGX), client-side (with Arm TrustZone) elements of the design, and the secure communication between them.
Our experimental evaluation shows that \oursystem provides full protection against data reconstruction, property inference, and membership inference attacks, whose outcomes are degraded to random guessing (\eg~white noise images or 50\% precision scores).
\oursystem is practical as it does not add significant overhead to the training process.
Compared to regular end-to-end \fl, \oursystem introduces a $3\times$ or higher delay for completing the training of all DNN layers. However, \oursystem achieves comparable ML performance when training only the first few layers, meaning that it is not needed to train all DNN layers. Due to this flexibility of layer-wise training, \oursystem can provide a similar ML model utility as end-to-end \fl, with fewer communication rounds ($0.54\times$), and a similar amount of network traffic ($1.002\times$), with only $\sim$15\% CPU time, $\sim$18\% memory usage, and $\sim$21\% energy consumption overhead at client-side.
\section{Background and Related Work}
\label{sec:related_work}

In this section, we provide the background needed to understand the way TEEs work (Sec.~\ref{subsec:trustedenvironments}), existing privacy risks in \fl (Sec.~\ref{subsec:privacyrisks}), privacy-preserving ML techniques using TEEs (Sec.~\ref{subsec:privacypreservingmltechniques}), as well as core ideas behind layer-wise \dnn training for \fl (Sec.~\ref{subsec:layerwise}).

\subsection{Trusted Execution Environments (TEE)}
\label{subsec:trustedenvironments}

A TEE enables the creation of a secure area on the main processor that provides strong confidentiality and integrity guarantees to any data and code it stores or processes.
TEEs realize strong isolation and attestation of secure compartments by enforcing a dual-world view where even compromised or malicious system (\ie~privileged) software in the normal world -- also known as the Rich Operating System Execution Environment (REE) -- cannot gain access to the secure world.
This allows for a drastic reduction of the TCB since only the code running in the secure world needs to be trusted. 
Another key aspect of TEEs is that they allow arbitrary code to run inside almost at native speed.
In order to keep the TCB as small as possible, current TEEs have limited memory; beyond this, TEEs are required to swap pages between secure and unprotected memory, which incurs a significant overhead and hence must be prevented.

Over the last few years, significant research and industry efforts have been devoted to developing secure and programmable TEEs for high-end devices (\eg~servers\footnote{Recently, cloud providers also offer TEE-enabled infrastructure-as-a-service solutions to their customers (\eg~Microsoft Azure Confidential).}) and mobile devices (\eg~smartphones).
In our work, we leverage Intel Software Guard Extensions (Intel SGX)~\cite{costan2016intel} at the server-side, while in the client devices we rely on Open Portable Trusted Execution Environment (OP-TEE)~\cite{optee2020}.
OP-TEE is a widely known open-source TEE framework that is supported by different boards equipped with Arm TrustZone. While some TEEs allow the creation of fixed-sized secure memory regions (e.g., of 128MB in Intel SGX), some others (e.g., ARM TrustZone) do not place any limit on the TEE size. However, creating large TEEs is considered to be bad practice since it has proven to significantly increase the attack surface. Therefore, the TEE size must always be kept as small as possible independently of the type of TEEs and devices being used. This principle has already been adopted by industry, \eg~in the HiKey 960 board the TEE size is only 16MiB.

%%%%%%%%%%%%%%%%%%%%%%%%%%%%%%%%%%%%
%%%%%%%%%%%% subsection %%%%%%%%%%%%
\subsection{Privacy Risks in \fl}
\label{subsec:privacyrisks}

Below we give a brief overview of the three main categories of privacy-related attacks in FL: data reconstruction, property inference, and membership inference attacks.

\mysubsubsection{Data Reconstruction Attack (DRA).}
The DRA aims at reconstructing original input data based on the observed model or its gradients.
It works by inverting model gradients based on generative adversarial attack-similar techniques~\cite{aono2017privacy, zhu2019deep, geiping2020inverting}, and consequently reconstructing the corresponding original data used to produce the gradients.
DRAs are effective when attacking \dnn's early layers, and when gradients have been only updated on a small batch of data (\ie~less than 8)~\cite{zhu2019deep, geiping2020inverting, mo2020darknetz}.
As the server typically observes updated models of each client in plaintext, it is more likely for this type of leakages to exist at the server.
By subtracting updated models with the global model, the server obtains gradients computed \wrt~clients' data during the local training. 

\mysubsubsection{Property Inference Attack (PIA).}
The goal of PIAs is to infer the value of private properties in the input data.
This attack is achieved by building a binary classifier trained on model gradients updated with auxiliary data and can be conducted on both server and client sides~\cite{melis2019exploiting}.
Specifically, property information, which also refers to the feature/latent information of the input data, is easier to be carried in stronger aggregation~\cite{mo2020layer}.
Even though clients in \fl only observe multiple snapshots of broadcast global models that have been linearly aggregated on participating clients' updates, property information can still be well preserved, providing attack points to client-side adversaries.

\mysubsubsection{Membership Inference Attack (MIA).}
The purpose of MIAs is to learn whether specific data instances are present in the training dataset.
One can follow a similar attack mechanism as PIAs to build a binary classifier when conducting MIAs~\cite{nasr2019comprehensive}, although there are other methods, \eg~using shadow models~\cite{shokri2017membership}.
The risk of MIAs can exist on both the server and client sides.
Moreover, because membership is `high-level' latent information, adversaries can perform MIAs on the final (well-trained) model and its last layer~\cite{nasr2019comprehensive, shokri2017membership, yeom2018privacy}.

%%%%%%%%%%%%%%%%%%%%%%%%%%%%%%%%%%%%
%%%%%%%%%%%% subsection %%%%%%%%%%%%
\subsection{Privacy-preserving ML using TEEs}
\label{subsec:privacypreservingmltechniques}

Running ML inside TEEs can hide model parameters from REE adversaries and consequently preserve privacy, as already used for light data analytics on servers~\cite{schuster2015vc3, ohrimenko2016oblivious} and for heavy computations such as \dnn training~\cite{hunt2018chiron, tramer2018slalom, gu2018securing, mo2020darknetz}.
However, due to TEEs' limited memory size, previous studies run only part of the model (\eg~sensitive layers) inside the TEE~\cite{tramer2018slalom, gu2018securing, mo2020darknetz, mo2019efficient}.
In the on-device training case, DarkneTZ~\cite{mo2020darknetz} runs the last layers with a Trusted Application inside TEEs to defend against MIAs, and leaves the first layers unprotected.
DarkneTZ's evaluation showed no more than 10\% overhead in CPU, memory, and energy on edge-like devices, demonstrating its suitability for client-side model updates in \fl.
In an orthogonal direction, several works leveraged clients' TEEs for verifying the integrity of local model training~\cite{chen2019deepattest, zhang2020enabling}, but did not consider privacy.
Considering a broader range of attacks (\eg~DRAs and PIAs), it is essential to protect all layers instead of the last layers only, something that \oursystem does.

%%%%%%%%%%%%%%%%%%%%%%%%%%%%%%%%%%%%
%%%%%%%%%%%% subsection %%%%%%%%%%%%
\subsection{Layer-wise DNN Training for \fl}
\label{subsec:layerwise}

Instead of training the complete \dnn model in an end-to-end fashion, one can train the model layer-by-layer from scratch, \ie~\emph{greedy layer-wise training}~\cite{bengio2006greedy, larochelle2009exploring}.
This method starts by training a shallow model (\eg~one layer) until its convergence.
Next, it appends one more layer to the converged model and trains only this new layer~\cite{belilovsky2019greedy}.
Usually, for each greedily added layer, the model developer builds a new classifier on top of it in order to output predictions and compute training loss.
Consequently, these classifiers provide multiple early exits, one per layer, during the forward pass in inference~\cite{kaya2019shallow}.
Furthermore, recently this method was shown to scale for large datasets such as ImageNet and to achieve performance comparable to regular end-to-end ML~\cite{belilovsky2019greedy}.
Notably, all previous studies on layer-wise training focused on generic ML.

\mysubsubsection{Contribution.}
Our work is the first to build a \dnn model in a \fl setting with privacy-preserving guarantees using TEEs, by leveraging the greedy layer-wise training, and to train each \dnn layer inside each \fl client's TEE.
Thus, \oursystem satisfies the constraint of TEE's limited memory while protecting the model from the aforementioned privacy attacks.
Interestingly, the classifiers built atop each layer may also provide personalization opportunities for the participating \fl clients.
\section{Threat Model and Assumptions}
\label{sec:threat_model}

\mysubsubsection{Threat model.}
We consider a standard \fl context where multiple client devices train a \dnn locally and send their (local) model parameters to a remote, centralized server, which aggregates these parameters to create a global model~\cite{mcmahan2017communication, bonawitz2019towards, kairouz2019advances}.
The goal of adversaries is to obtain sensitive information embedded in the global model through data reconstruction~\cite{zhu2019deep, geiping2020inverting} or inference attacks~\cite{melis2019exploiting, nasr2019comprehensive}.
We consider two types of (passive) adversaries: (i) users of client devices who have access to distinct snapshots of the global model and (ii) the server's owner (\eg~a cloud or edge provider) who has access to the updated model gradients.
Adversaries are assumed to be \emph{honest-but-curious}, meaning that they allow \fl algorithms to run as intended while trying to infer as much information as possible from the global model or gradients.
Adversaries can have full control (\ie~root privileges) of the server or the client device, and can perform their attacks against any \dnn layer.
However, attacks against the TEE, such as side-channel attacks (\eg~Voltpillager~\cite{Chen2021voltpillager}), physical attacks (\eg~Platypus~\cite{Lipp2021Platypus}) and those that exploit weaknesses in TEEs (\eg~\cite{10.5555/3241189.3241231}) and their SDKs (\eg~\cite{10.1145/3319535.3363206}) are out of scope for this paper.

\mysubsubsection{Assumptions.}
We assume that the server and enough participating \fl client devices have a TEE whose memory size is larger than the largest layer of the \dnn to be trained.
This is the case in current \fl \dnns.
However, in the unlikely case that a layer does not fit in available TEEs, the network design needs to be adjusted with smaller, but more layer(s), or a smaller training batch size.
We also assume that there is a secure way to bootstrap trust between the server TEE and each of the client device TEE (\eg~using a slightly modified version of the SIGMA key exchange protocol~\cite{krawczyk2003sigma, zhao2019sectee}, or attested TLS~\cite{knauth2018integrating}), and that key management mechanisms exist to update and revoke keys when needed~\cite{trustanchors}. Finally, we assume that the centralized server will forward data to/from its TEE. Yet, it is important to note that if the server was malicious and would not do this, this would only affect the availability of the system (i.e., the security and privacy properties of our solution remain intact). This type of Denial-of-Service (DoS) attack is hard to defend against and is not considered within the standard TEE threat model.
\section{\oursystem Framework}
\label{sec:framework}

%%%%%%%%%%%% overview figure %%%%%%%%%%%%
\begin{figure*}[ht]
    \centering
    \includegraphics[width=2\columnwidth]{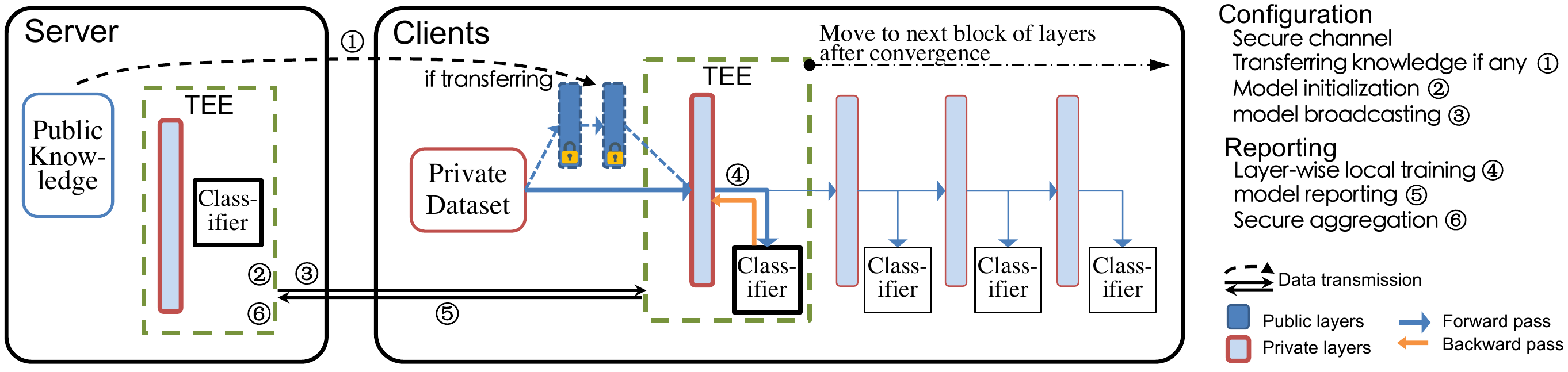}
    \caption{A schematic diagram of the \oursystem{} framework. The main phases follow the system design in~\cite{bonawitz2019towards}.}
    \label{fig:framework_overview}
\end{figure*}

In this section, we first present an overview of the proposed system and its functionalities (Sec.~\ref{sec:system-overview}), and then detail how the framework employs layer-wise training and aggregation in conjunction to TEEs in \fl (Sec.~\ref{sec:layer-wise-training-in-tees}).

%%%%%%%%%%%%%%%%%%%%%%%%%%%%%%%%%%%%
%%%%%%%%%%%% subsection %%%%%%%%%%%%
\subsection{System Overview}
\label{sec:system-overview}

We propose a \oursystemlong framework which allows clients to collaboratively train a DNN model while keeping the model's layers always inside TEEs during training.
Figure~\ref{fig:framework_overview} provides an overview of the framework and the various steps of the \emph{greedy layer-wise training and aggregation}.
In general, starting from the first layer, each layer is trained until convergence, before moving to the next layer.
In this way, \oursystem aims to achieve full privacy preservation without significantly increasing system cost.
\oursystem's design provides the following functionalities:

\mysubsubsection{Privacy-by-design Guarantee.}
\oursystem ensures that layers are always protected from adversaries while they are being updated.
Privacy risks depend on the aggregation level and frequency with which they happen, when exposing the model or its layers~\cite{melis2019exploiting, geiping2020inverting, mo2020layer}.
In \oursystem, lower-level information (\ie~original data and attributes) is not exposed because updated gradients during training are not accessible from adversaries (they happen inside the TEEs). This protects against DRAs and PIAs.  
However, when one of such layers is exposed after convergence, there is a risk of MIAs.
We follow a more practical approach based on the observation that membership-related information is only sensitive in the \emph{last DNN layer}, making it vulnerable to MIAs as indicated in previous research~\cite{nasr2019comprehensive, sablayrolles2019white, mo2020darknetz,mo2020layer}.
To avoid this risk on the final model, \oursystem can keep the last layer inside the clients TEEs after training.

\mysubsubsection{Device Selection.} After the server and a set of TEE-enabled clients agree on the training of a DNN model via FL, clients inform the server about their TEE's memory constraints. The server then (re)constructs a DNN model suitable for this set of clients and selects the clients that can accommodate the model layers within their TEE. In each round, the server can select new clients and the device selection algorithm can follow existing FL approaches~\cite{nishio2019client, huang2020efficiency}.

\mysubsubsection{Secure Communication Channels.} The server establishes two secure communication channels with each of its clients: (i) one from its REE to the client's REE (\eg~using TLS) to exchange data with clients and (ii) a logical one from its TEE to the client's TEE for securely exchanging private information (\eg~model layer training information).
In the latter case, the transmitted data is encrypted using cryptographic keys known only to the server and client TEEs and is sent over the REE-REE channel. 
It is important to note that the secure REE-REE channel is only an additional security layer.
All privacy guarantees offered by \oursystem are based on the hardware-backed cryptographic keys stored inside TEEs.

\mysubsubsection{Model Initialization and Configuration.}
The server configures the model architecture, decides the layers to be protected by TEEs, and then initializes model parameters inside the TEE (step~\textcircled{\raisebox{-.9pt} {2}}, Fig.~\ref{fig:framework_overview}).
The latter ensures clients' local training starts with the same weight distribution~\cite{mcmahan2017communication, wang2020federated}.
In addition, the server configures other training hyper-parameters such as learning rate, batch size, and epochs, before transmitting such settings to the clients (step~\textcircled{\raisebox{-.9pt} {3}}, Fig.~\ref{fig:framework_overview}).

In cases of typical ML tasks such as image recognition where public knowledge is available such as pre-trained \dnn models or public datasets with features similar to the client private data, the server can transfer this knowledge (especially in cross-device \fl~\cite{kairouz2019advances}) in order to bootstrap and speed up the training process.
In both cases, this knowledge is contained in the first layers.
Thus, the clients leave the first layers frozen and only train the last several layers of the global model.
This training process is similar to the concept of transfer learning~\cite{pan2009survey,brownlee2019gentle,torrey2010transfer}, where, in our case, public knowledge is transferred in a federated manner.

In \oursystem, the server can learn from \textit{public models}. 
Thus, during initialization, the server first chooses a model pre-trained on public data that have a similar distribution with private data. 
The server keeps the first layers, removes the last layer(s), and assembles new layer(s) atop the reserved first ones.
These first layers are transferred to clients and are always kept frozen (step~\textcircled{\raisebox{-.9pt} {1}}, Fig.~\ref{fig:framework_overview}).
New layers, attached to the reserved layers, are trained inside each client's TEE, and then aggregated inside the server's TEE (steps~\textcircled{\raisebox{-.9pt} {2}}$\sim$\textcircled{\raisebox{-.9pt} {6}}, Fig.~\ref{fig:framework_overview}).
In learning from \textit{public datasets}, the server first performs an initial training to build the model based on public datasets.

\mysubsubsection{Local Training.}
After model transmission and configuration using secure channels, each client starts local training on their data on each layer via a model partitioned execution technique (step~\textcircled{\raisebox{-.9pt} {4}}, Fig.~\ref{fig:framework_overview}). 
We detail this step in Sec.~\ref{sec:layer-wise-training-in-tees}.

\mysubsubsection{Reporting and Aggregation.}
Once local training of a layer is completed inside TEEs, all participating clients report the layer parameters to the server through secure channels (step~\textcircled{\raisebox{-.9pt} {5}}, Fig.~\ref{fig:framework_overview}).
Finally, the server securely aggregates the received parameters within its TEE and applies FedAvg~\cite{mcmahan2017communication}, resulting in a new global model layer (step \textcircled{\raisebox{-.9pt} {6}}, Fig.~\ref{fig:framework_overview}).

%%%%%%%%%%%%%%%%%%%%%%%%%%%%%%%%%%%%
%%%%%%%%%%%% subsection %%%%%%%%%%%%
\subsection{Layer-wise Training and Aggregation}
\label{sec:layer-wise-training-in-tees}

In order to address the problem of limited memory inside a TEE when training a \dnn model, we modify the greedy layer-wise learning technique proposed in~\cite{bengio2006greedy} for general \dnn training~\cite{belilovsky2019greedy}, to work in the \fl setting.
The procedure of layer-wise training and aggregation is detailed in the following Algorithms~\ref{algo:ppfl_server} and~\ref{algo:ppfl_client}.

\mysubsubsection{Algorithm~\ref{algo:ppfl_server}.}
This algorithm details the actions taken by \oursystem on the server side.
When not specified, operations are carried out outside the TEE (\ie~in the REE).
First, the server initializes the global \dnn model with random weights or public knowledge (steps~\textcircled{\raisebox{-.9pt} {1}}-\textcircled{\raisebox{-.9pt} {2}}, Fig.~\ref{fig:framework_overview}).
Thus, each layer $l$ to be trained is initialized ($\bm{\theta}_{l}$) and prepared for broadcast.
The server checks all available devices and constructs a set of participating clients whose TEE is larger than the required memory usage of $l$.
Then, it broadcasts the model's layer to these participating clients (step \textcircled{\raisebox{-.9pt} {3}}, Fig.~\ref{fig:framework_overview}), via \texttt{ClientUpdate()} (see Algorithm~\ref{algo:ppfl_client}).
Upon receiving updates from all participating clients, the server decrypts the layer weights, performs secure layer aggregation and averaging inside its TEE (step \textcircled{\raisebox{-.9pt} {6}}), and broadcasts the new version of $l$ to the clients for the next \fl round.
Steps~\textcircled{\raisebox{-.9pt} {2}}$\sim$\textcircled{\raisebox{-.9pt} {6}} are repeated until the training of $l$ converges, or a fixed number of rounds are completed.
Then, this layer is considered fully trained ($\bm{\theta}_{l}^0$), it is passed to the REE, and is broadcasted to all clients to be used for training the next layer.
Interestingly, \oursystem also allows grouping \emph{multiple layers into blocks} and training each block inside client TEEs in a similar fashion as the individual layers.
This option allows for better utilization of the memory space available inside each TEE and reduces communication rounds for the convergence of more than one layer at the same time.

\mysubsubsection{Algorithm~\ref{algo:ppfl_client}}.
This algorithm details the actions taken by \oursystem on the client side.
Clients load the received model parameters from the server and decrypt and load the target training layer $l$ inside their TEEs.
More specifically, in the front, this new layer $l$ connects to the previous pre-trained layer(s) that are frozen during training.
In the back, the clients attach on $l$ their \emph{own derived classifier}, which consists of fully connected layers and a softmax layer as the model exit.
Then, for each epoch, the training process iteratively goes through batches of data and performs both \emph{forward and backward passes}~\cite{lecun2015deep} to update both the layer under training and the classifier inside the TEE (step~\textcircled{\raisebox{-.9pt} {4}}, Fig.~\ref{fig:framework_overview}).
During this process, a \emph{model partitioned execution} technique is utilized, where intermediate representations of the previously trained layers are passed from the REE to the TEE via shared memory in the forward pass. 
After local training is completed (\ie~all batches and epochs are done), each client sends via the secure channel the (encrypted) layer's weights from its TEE to the server's TEE (step~\textcircled{\raisebox{-.9pt} {5}}).

%%%%%%%%%%%% algorithm 1 %%%%%%%%%%%%
\begin{algorithm}[t!]
\SetAlgoLined
    \textbf{Input:}\\
    \begin{itemize}[leftmargin=15pt]
        \item Number of all clients: $N$
        \item TEE memory size of Client $n$: $S^{(n)}$
        \item Memory usage of layers $\{1,...,L\}$ in training (forward and backward pass in total): $\{S_{1},...,S_{L}\}$
        \item Communication rounds: $R$
    \end{itemize}
    \textbf{Output:} Aggregated final parameters: $\{\bm{\theta}_1^0, ..., \bm{\theta}_L^0\}$\\
    \vspace{5pt}
    \% \emph{Layer-wise client updates} \\
    \For{$l \in \{1,...,L\}$}{
        \vspace{5pt}
        \% \emph{Select clients with enough TEE memory}\\
        Initialize participating client list $\mathbf{J} = \{\}$ \\
        \For{$n \in \{1, ..., N\}$}{
            \If{$S^{(n)}$ > $S_l$}{
                $\mathbf{J} \xleftarrow{} \mathbf{J} \cup \{n\}$
            }
        }
        \vspace{5pt}
        Initialize $\bm{\theta}_{l}$ (parameters of layers $l$)\ \textcolor{forestgreen}{in TEE}\\
        \vspace{5pt}
        \For{$r \in \{1,...,R\}$}{
            \For{$j \in \mathbf{J}$}{
                \vspace{3pt}
                \% \emph{clients' local updating: see Algorithm 2} \\ 
                 $\bm{\theta}_{l}^{(j)}$=\textbf{ClientUpdate}($l,\bm{\theta}_{l}$)
            }
            \vspace{5pt}
            \% \emph{FedAvg with Secure Aggregation} \\
            $\bm{\theta}_{l} = \frac{1}{\mathrm{size}(\mathbf{J})} \sum_{j\in \mathbf{J}}\bm{\theta}_{l}^{(j)}$ \textcolor{forestgreen}{in TEE}\\
        }
        \vspace{5pt}
        Save $\bm{\theta}_{l}$ from TEE as $\bm{\theta}_{l}^0$ in REE
    }
    \vspace{5pt}
    return $\{\bm{\theta}_1^0, ..., \bm{\theta}_L^0\}$
    \caption{\textbf{\oursystem-Server with TEE}}
    \label{algo:ppfl_server}
\end{algorithm}

%%%%%%%%%%%% algorithm 2 %%%%%%%%%%%%
\begin{algorithm}[t!]
\SetAlgoLined
    \textbf{Initialization:}
    \begin{itemize}[leftmargin=15pt]
        \item Local dataset $\mathcal{X}$: data $\{\bm{x}\}$ and labels $\{\bm{y}\}$
        \item Trained final parameters of all previous layers, \ie~$\bm{\theta}_1^0, \bm{\theta}_2^0,...,\bm{\theta}_{l-1}^0$
        \item Number of local training epochs: $E$
        \item Activation function: $\sigma()$ and loss function: $\ell$
        \item Classifier: $C()$
    \end{itemize}
    \textbf{Input:}
    \begin{itemize}[leftmargin=15pt]
        \item Target layer: $l$
        \item Broadcast parameters of layer $l$: $\bm{\theta}_{l}$
    \end{itemize}
    \textbf{Output:} Updated parameters of layer $l$: $\bm{\theta}_{l}$\\
    \vspace{5pt}
    \% \emph{Weights and biases of layers $1,...,(l-1)$ and $l$} \\
    \For{$i \in \{1,...,l-1\}$}{
        $\{\bm{W}_{i}, \bm{b}_{i}\} \xleftarrow{} \bm{\theta}_i^0$
    }
    $\{\bm{W}_{l}, \bm{b}_{l}\} \xleftarrow{} \bm{\theta}_{l}$ \textcolor{forestgreen}{in TEE}\\
    \vspace{5pt}
    \% \emph{Training process} \\
    \For{$e \in \{1,..,E\}$}{
        \For{$\{\bm{x},\bm{y}\} \in \mathcal{X}$}{
            \vspace{5pt}
            \% \emph{Forward pass}\\
            Intermediate representation $\bm{T}_{0}=\bm{x}$\tikzmark{right}\\
            \For{$i \in \{1,...,l-1\}$}{
                $\bm{T}_i = \sigma(\bm{W}_i \bm{T}_{i-1} + \bm{b}_i)$\\
            }
            \vspace{5pt}
            %\For{$l \in K$}{
            $\bm{T}_l = \sigma(\bm{W}_l \bm{T}_{l-1} + \bm{b}_l)$\tikzmark{top}\\
            $\ell \xleftarrow{} \ell(C(\bm{T}_{l}), \bm{y})$\\
            \vspace{5pt}
            \% \emph{Backward pass}\\
            $\frac{\partial \ell}{\partial C}$ to update parameters of $C$\\
            \vspace{3pt}
            \% \emph{Updating layer l}\\
            $\bm{W}_{l} \xleftarrow{} \bm{W}_{l} + \frac{\partial \ell}{\partial \bm{W}_l}$; $\bm{b}_{l} \xleftarrow{} \bm{b}_{l} + \frac{\partial \ell}{\partial \bm{b}_l}$ \tikzmark{bottom}
        }
    }
    \vspace{3pt}
    $\bm{\theta}_{l} = \{ \bm{W}_{l}, \bm{b}_{l}\}$ \textcolor{forestgreen}{in TEE}\\
    \vspace{5pt}
    return $\bm{\theta}_{l}$
    \caption{\textbf{ClientUpdate}$(l, \bm{\theta}_{l})$ with TEEs}
    \label{algo:ppfl_client}
    \AddNote{top}{bottom}{right}{in TEE}
\end{algorithm}

%%%%%%%%%%%%%%%%%%%%%%%%%%%%%%%%%%%%
%%%%%%%%%%%% subsection %%%%%%%%%%%%
\mysubsubsection{Model Partitioned Execution.}
The above learning process is based on a technique that conducts model training (including both forward and backward passes) across REEs and TEEs, namely model partitioned execution.
The transmission of the forward activations (\ie~intermediate representation) and updated parameters happens between the REE and the TEE via \emph{shared memory}.
On a high level, when a set of layers is in the TEE, activations are transferred from the REE to the TEE (see Algorithm~\ref{algo:ppfl_client}). Assuming global layer $l$ is under training, the layer with its classifier $C(.)$ are executed in the TEE, and the previous layers (\ie~$1$ to $l-1$) are in the REE.

Before training, layer $l$'s parameters are loaded and decrypted securely within the TEE.
During the \emph{forward pass}, local data $\bm{x}$ are inputted, and the REE processes the previous layers from $1$ to $l-1$ and invokes a command to transfer the layer $l-1$'s activations (\ie~$T_{l-1}$) to the secure memory through a buffer in shared memory.
The TEE switches to the corresponding invoked command in order to receive layer $l-1$'s activations and processes the forward pass of layer $l$ and classifier $C(.)$ in the TEE.

During the \emph{backward pass}, the TEE computes the $C(.)$'s gradients based on received labels $\bm{y}$ and outputs of $C(.)$ (produced in the forward pass) and uses them to compute the gradients of the layer $l$ in the TEE.
The training of this batch of data (\ie~$\bm{x}$) finishes here, and there is no need to transfer $l$'s errors from the TEE to the REE via shared memory, as previous layers are frozen outside the TEE.
After that, the parameters of layer $l$ are encrypted and passed to the REE, ready to be uploaded to the server, corresponding to the $FedSGD$~\cite{chen2016revisiting}.
Further, $FedAvg$~\cite{mcmahan2017communication} which requires multiple batches to be processed before updating, repeats the same number of forward and backward passes across the REE and the TEE for each batch of data.

\mysubsubsection{Algorithmic Complexity Analysis.}
Next, we analyze the algorithmic complexity of \oursystem and compare it to standard end-to-end \fl.
For the global model's layers $l \in \{1,\dots,L\}$, we denote the forward and backward pass cost on layer $l$ as $\mathtt{F}_l$ and $\mathtt{B}_l$, respectively. 
The corresponding cost on the classifier is denoted as $\mathtt{F}_c$ and $\mathtt{B}_c$.
Then, in end-to-end \fl, the total training cost for one client is:
\begin{equation}
    \left(\sum_{l=1}^L (\mathtt{F}_l + \mathtt{B}_l) + \mathtt{F}_c + \mathtt{B}_c\right) \cdot S \cdot E
\label{eq:fbpass_e2e}
\end{equation}
\noindent where $S$ is the number of steps in one epoch (\ie~number of samples inside local datasets divided by the batch size).
As in \oursystem all layers before the training layer $l$ are kept frozen, the cost of training layer $l$ is $(\sum_{k=1}^l \mathtt{F}_k + \mathtt{F}_c + \mathtt{B}_l + \mathtt{B}_c) \cdot S \cdot E$.
Then, by summation, we get the total cost of all layers as:
\begin{equation}
    \left(\sum_{l=1}^L \sum_{k=1}^l \mathtt{F}_k + \sum_{l=1}^L \mathtt{B}_l + L \cdot (\mathtt{F}_c + \mathtt{B}_c) \right) \cdot S \cdot E
\label{eq:fbpass_ppfl}
\end{equation}
By comparing Equations~\ref{eq:fbpass_e2e} and~\ref{eq:fbpass_ppfl}, we see the overhead of \oursystem comes from: (i) repeated forward pass in previous layers ($l \in \{1,\dots, l-1\}$) when training layer $l$, and (ii) repeated forward and backward pass for the classifier atop layer $l$.
\section{Implementation \& Evaluation Setup}

In this section, we first describe the implementation of the \oursystem system (Sec.~\ref{sec:implementation}), and then detail how we assess its performance on various \dnn models and datasets (Sec.~\ref{sec:models-datasets}) using different metrics (Sec.~\ref{sec:performance-metrics}).
We follow common setups of past \fl systems~\cite{mcmahan2017communication, wang2020federated} and on-device TEE works~\cite{mo2020darknetz, amacher2019performance}. 

%%%%%%%%%%%%%%%%%%%%%%%%%%%%%%%%%%%%
%%%%%%%%%%%% subsection %%%%%%%%%%%%
\subsection{\oursystem Prototype}
\label{sec:implementation}

We implement the \emph{client-side} of \oursystem by building on top of DarkneTZ~\cite{mo2020darknetz}, 
in order to support on-device \fl with Arm TrustZone.
In total, we changed 4075 lines of code of DarkneTZ in C.
We run the client-side on a HiKey 960 Board, which has four ARM Cortex-A73 and four ARM Cortex-A53 cores configured at 2362MHz and 533MHz, respectively, as well as a 4GB LPDDR4 RAM with 16MiB TEE secure memory (\ie~TrustZone).
Since the CPU power/frequency setting can impact the TrustZone's performance~\cite{amacher2019performance}, we execute the on-device \fl training with full CPU frequency.
In order to emulate multiple device clients and their participation in \fl rounds, we use the HiKey board in a repeated, iterative fashion, one time per client device.
We implement the \emph{server-side} of \oursystem on generic Darknet ML framework~\cite{darknet13} by adding 751 lines of C code based on Microsoft OpenEnclave~\cite{microsoft2020oe} with Intel SGX.
For this, an Intel Next Unit of Computing (ver.NUC8BEK, i3-8109U CPU, 8GB DDR4-2400MHz) was used with SGX-enabled capabilities.

Besides, we developed a set of bash shell scripts to control the \emph{\fl process} and create the \emph{communication channels}.
For the communication channels between server and client to be secure, we employ standard cryptographic-based network protocols such as \texttt{SSH} and \texttt{SCP}.
All data leaving the TEE are encrypted using the Advanced Encryption Standard (AES) in Cipher Block Chaining (CBC) mode with random Initialization Values (IV) and 128-bit cryptographic keys.
Without loss of generality, we opted for manually hardcoding the cryptographic keys inside the TEEs ourselves.
Despite key management in TEE-to-TEE channels being an interesting research problem, we argue that establishing, updating, and revoking keys do not happen frequently and hence the overhead these tasks introduce is negligible compared to one from the \dnn training.

The implementation of \oursystem server and client is available for replication and extension: \url{https://github.com/mofanv/PPFL}.

%%%%%%%%%%%%%%%%%%%%%%%%%%%%%%%%%%%%
%%%%%%%%%%%% subsection %%%%%%%%%%%%
\subsection{Models and Datasets}
\label{sec:models-datasets}

We focus on Convolutional Neural Networks (CNNs) since the privacy risks we consider (Sec.~\ref{sec:threat_model} and~\ref{sec:system-overview}) have been extensively studied on such \dnns~\cite{melis2019exploiting, nasr2019comprehensive}.
Also, layer-based learning methods mostly aim at CNN-like \dnns~\cite{belilovsky2019greedy}.
Specifically, in our \oursystem evaluation, we employ \dnns commonly used in the relevant literature (Table~\ref{tab:dnns-evaluation}).

\begin{table}[t!]
\small
\setlength\tabcolsep{3pt}
\caption{\dnns used in the evaluation of \oursystem. }
\label{tab:dnns-evaluation}
\begin{tabular}{ll}
\hline
DNN & Architecture\\
\hline
LeNet~\cite{lecun1998gradient,mcmahan2017communication} & C20-MP-C50-MP-FC500-FC10 \\
\hline
AlexNet~\cite{krizhevsky2017imagenet, belilovsky2019greedy} & C128$\times$3-AP16-FC10 \\
\hline
VGG9~\cite{simonyan2014very, wang2020federated} & \vtop{\hbox{\strut C32-C64-MP-C128$\times$2-MP-D0.05-C256$\times$2} \hbox{\strut -MP-D0.1-FC512$\times$2-FC10}} \\
\hline
VGG16~\cite{simonyan2014very} & \vtop{\hbox{\strut C64$\times$2-MP-C128$\times$2-MP-C256$\times$3-C512$\times$3} \hbox{\strut -MP-FC4096$\times$2-FC1000-FC10}} \\
\hline
MobileNetv2~\cite{sandler2018mobilenetv2}& 68 layers, unmodified refer to~\cite{sandler2018mobilenetv2} for details \\
\hline
\end{tabular}
\begin{flushleft}
\footnotesize{
Architecture notation: Convolution layer (C) with a given number of filters; filter size is $5\times5$ in LeNet and $3\times3$ in AlexNet, VGG9, and VGG16. Fully Connected (FC) with a given number of neurons. All C and FC layers are followed by ReLU activation functions. MaxPooling (MP). AveragePooling (AP) with a given stride size. Dropout layer (D) with a given dropping rate.} \hfill \\
\end{flushleft}
\vspace{-5pt}
\end{table}

For our experimental analysis, we used $MNIST$ and $CIFAR10$, two datasets commonly employed by \fl researchers.
Note that in practice, \fl training needs labeled data locally stored at the clients' side.
Indeed, the number of labeled examples expected to be present in a real setting could be fewer than what these datasets may allocate per \fl client.
Nonetheless, using them allows comparison of our results with state-of-art end-to-end \fl methods~\cite{li2018federated, wang2020federated, geiping2020inverting}.

Specifically, LeNet is tested on MNIST~\cite{lecun1998gradient} and all other models are tested on CIFAR10~\cite{krizhevsky2009learning}.
The former is a handwritten digit image ($28$$\times$$28$) dataset consisting of $60k$ training samples and $10k$ test samples with 10 classes.
The latter is an object image ($32$$\times$$32$$\times$$3$) dataset consisting of $50k$ training samples and $10k$ test samples with 10 classes.
We follow the setup in~\cite{mcmahan2017communication} to partition training datasets into 100 parts, one per client, in two versions: i) Independent and Identically Distributed (IID) where a client has samples of all classes; ii) Non-Independent and Identically Distributed (Non-IID) where a client has samples only from two random classes.

%%%%%%%%%%%%%%%%%%%%%%%%%%%%%%%%%%%%
%%%%%%%%%%%% subsection %%%%%%%%%%%%
\subsection{Performance Metrics}
\label{sec:performance-metrics}

The evaluation of \oursystem prototype presented in the next section focuses on assessing the framework from the point of view of (i) privacy of data, (ii) ML model performance,  and (iii) client-side system cost.
Although ML computations (\ie~model training) have the same precision and accuracy no matter in REEs or TEEs, \oursystem changes the \fl model training process into a layer-based training.
This affects ML accuracy and the number of communication rounds needed for the model to converge (among others).
Thus, we devise several metrics and perform extensive measurements to assess overall \oursystem performance.
We conduct system cost measurements only on client devices since their computational resources are more limited compared to the server.
All experiments are done with 10\% of the total number of clients (\ie~10 out of 100) participating in each communication round. We run FL experiments on our \oursystem prototype (Sec.~\ref{sec:implementation}) to measure the system cost. To measure privacy risks and ML model performance, we perform simulations on a cluster with multiple NVIDIA RTX6000 GPUs (24GB) nodes running PyTorch v1.4.0 under Python v3.6.0.

\mysubsubsection{Model Performance.}
We measure three metrics to assess the performance of the model and \oursystem-related process:
\begin{enumerate}[leftmargin=12pt]
    \item \emph{Test Accuracy}: ML accuracy of test data on a given \fl model, for a fixed number of communication rounds.
    \item \emph{Communication Rounds}: Iterations of communication between server and clients needed to achieve a particular test accuracy.
    \item \emph{Amount of communication}: Total amount of data exchanged to reach a test accuracy. Transmitted data sizes may be different among communication rounds when considering different layers' sizes in layer-wise training.
\end{enumerate}

\mysubsubsection{Privacy Assessment.}
We measure privacy risk of \oursystem by applying three \fl-applicable, privacy-related attacks:
\begin{enumerate}[leftmargin=14pt]
    \item Data Reconstruction Attack (DRA)~\cite{zhu2019deep}
    \item Property Inference Attack (PIA)~\cite{melis2019exploiting}
    \item Membership Inference Attack (MIA)~\cite{nasr2019comprehensive}
\end{enumerate}
We follow the proposing papers and their settings to conduct each attack on the model trained in \fl process.

\mysubsubsection{Client-side System Cost.}
We monitor the efficiency of client on-device training, and measure the following device costs for \oursystem-related process information:
\begin{enumerate}[leftmargin=12pt]
    \item \emph{CPU Execution Time (s)}: Time the CPU was used for processing the on-device model training, including time spent in REE and the TEE's user and kernel time, which is reported by using function \texttt{getrusage(RUSAGE\_SELF)}.
    \item \emph{Memory Usage (MB)}:
    We add REE memory (the maximum resident set size in RAM, accessible by \texttt{getrusage()}) and allocated TEE memory (accessible by \texttt{mdbg\_check(1)}) to get the total memory usage.
    \item \emph{Energy Consumption (J)}: 
    Measured by all energy used to perform one on-device training step when the model runs with/without TEEs. For this, we use the \emph{Monsoon High Voltage Power Monitor}~\cite{msoon}. We configure the power to HiKey board as 12V voltage while recording the current in a $50Hz$ sampling rate.
    Training with a high-performance power setting can lead to high temperature and consequently under-clocking. Thus, we run each trial with 2000 steps continuously, starting with 120s cooling time.
\end{enumerate}

\section{Evaluation Results}

In this section, we present the experimental evaluation of \oursystem aiming to answer a set of key questions.

%%%%%%%%%%%%%%%%%%%%%%%%%%%%%%%%%%%%
%%%%%%%%%%%% subsection %%%%%%%%%%%%
\subsection{How Effectively does \oursystem Thwart Known Privacy-related Attacks?}
\label{sec:privacy-results}

To measure the exposure of the model to known privacy risks, we conduct data reconstruction, property inference, and membership inference attacks (\ie~DRAs, PIAs, and MIAs) on the \oursystem model.
While training AlexNet and VGG9 models on CIFAR10 in an IID setting.
We compare the exposure of \oursystem to these attacks against a standard, end-to-end \fl-trained model.
Table~\ref{tab:privacy_results} shows the average performance of each attack in the same way it is measured in literature~\cite{zhu2019deep, melis2019exploiting, nasr2019comprehensive}: Mean-Square-Error (MSE) for the DRA, Area-Under-Curve (AUC) for the PIA, and Precision for the MIA.

From the results, it becomes clear that, while these attacks can successfully disclose private information in regular end-to-end \fl, they fail in \oursystem.
As DRAs and PIAs rely on intermediate training models (\ie~gradients) that remain protected, \oursystem can fully defend against them. The DRA can only reconstruct a fully noised image for any target image (\ie~an MSE of $\sim$1.3 for the specific dataset), while the PIA always reports a random guess on private properties (\ie~an AUC of $\sim$0.5).
Regarding the MIA on final trained models, as \oursystem keeps the last layer and its outputs always protected inside the client's TEE, it forces the adversary to access only previous layers, which significantly drops the MIA's advantage (\ie~Precision$\approx$0.5).
Thus, \oursystem fully addresses privacy issues raised by such attacks.

\begin{table}[t!]
\small
\caption{Results of three privacy-related attacks (DRA, PIA and MIA) on \oursystem vs. end-to-end (E2E) \fl.
Average score reported with 95\% confidence interval in parenthesis.}
\label{tab:privacy_results}
\setlength\tabcolsep{2pt}
\begin{tabular}{lllll}
\hline
\multicolumn{1}{c}{\rule{0pt}{3ex} \multirow{2}{*}{\begin{tabular}[c]{@{}c@{}}Learning\\ Method\end{tabular}}} & \multicolumn{1}{c}{\multirow{2}{*}{Model}}             & \multicolumn{3}{c}{Privacy-related Attack}                                                                                                                                                                                            \\ \cline{3-5} 
\multicolumn{1}{c}{}                                                                           & \multicolumn{1}{c}{}                                   & \multicolumn{1}{c}{\rule{0pt}{3ex} DRA\begin{scriptsize}, in MSE \end{scriptsize}$^{\alpha}$}                                       & \multicolumn{1}{c}{PIA\begin{scriptsize}, in AUC \end{scriptsize}$^{\delta}$}                                       & \multicolumn{1}{c}{MIA\begin{scriptsize}, in Prec. \end{scriptsize}$^{\epsilon}$}                                      \\ \hline
\rule{0pt}{4ex} E2E     & \begin{tabular}[c]{@{}l@{}}AlexNet\\ VGG9\end{tabular} & \raisebox{-5pt}{\includegraphics[width=0.035\textwidth]{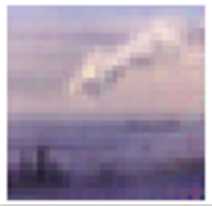}} \begin{tabular}[c]{@{}l@{}}0.017 (0.01)\\ 0.008 (<0.01)\end{tabular} & \begin{tabular}[c]{@{}l@{}}0.930 (0.03)\\ 0.862 (0.05)\end{tabular} & \begin{tabular}[c]{@{}l@{}}0.874 (0.01)\\ 0.765 (0.04)\end{tabular} \\ \hline
\rule{0pt}{4ex} PPFL                                                                                           & \begin{tabular}[c]{@{}l@{}}AlexNet\\ VGG9\end{tabular} & \raisebox{-5pt}{\includegraphics[width=0.035\textwidth]{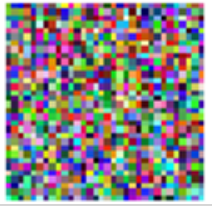}} $\sim$1.3                                                             & $\sim$0.5                                                             & \begin{tabular}[c]{@{}l@{}}0.506 (0.01)\\ 0.507 (<0.01)\end{tabular} \\ \hline
\end{tabular}
\begin{flushleft}
\footnotesize{
$^\alpha$MSE (mean-square error) measures the difference between constructed images and target images (range is $[0, \infty)$, and the lower MSE is, the more privacy loss);
$^\delta$AUC refers to the area under receiver operating curve;
$^\epsilon$Prec. refers to Precision.
The range of both AUC and Prec. is $[0.5, 1]$ (assuming 0.5 is for random guesses), and the higher AUC or Prec. is, the more privacy loss).}
\end{flushleft}
\end{table}

%%%%%%%%%%%%%%%%%%%%%%%%%%%%%%%%%%%%
%%%%%%%%%%%% subsection %%%%%%%%%%%%
\subsection{What is the \oursystem Communication Cost?}
\label{sec:communicationcost-results}

\mysubsubsection{Predefined ML Performance.}
Next, we measure \oursystem's communication cost to complete the \fl process, when a specific ML performance is desired.
For this, we first execute the standard end-to-end \fl without TEEs for 150 rounds and record the achieved ML performance.
Subsequently, we set the same test accuracy as a requirement, and measure the number of communication rounds and amount of communication required by \oursystem to achieve this ML performance.

In this experiment, we set the number of local epochs at clients as 10.
We use SGD as the optimization algorithm and set the learning rate as 0.01, with a decay of 0.99 after each epoch.
Momentum is set to 0.5 and the batch size to 16.
When training each layer locally, we build one classifier on top of it.
The classifier's architecture follows the last convolutional (Conv) layer and fully-connected (FC) layers of the target model (\eg~AlexNet or VGG9).
Thus, the training of each global model's layer progresses until all Conv layers are finished.
We choose AlexNet and VGG9 on CIFAR10, because MNIST is too simple for testing.
Then, the classifier atop all Conv layers is finally trained to provide outputs for the global model.
Note that we also aggregate the client classifiers while training one global layer to provide the test accuracy after each communication round.
We perform these experiments on IID and Non-IID data.

Overall, the results in Table~\ref{tab:overal_performance} show that, while trying to reach the ML performance achieved by the standard end-to-end \fl system, \oursystem adds small communication overhead, if any, to the \fl process.
In fact, in some cases, it can even reduce the communication cost, while preserving privacy when using TEEs.
As expected, using Non-IID data leads to lower ML performance across the system, which also implies less communication cost for \oursystem as well.

The reason why in many cases \oursystem has reduced communication cost, while still achieving comparable ML performance, is that training these models on datasets such as CIFAR10 may not require training the complete model.
Instead, during the early stage of \oursystem's layer-wise training (\eg~first global layer+classifier), it can already reach good ML performance, and in some cases even better than training the entire model.
We explore this aspect further in the next subsection.
Consequently, and due to the needed rounds being fewer, the amount of communication is also reduced.

The increased cost when training VGG9 is due to the large number of neurons in the classifier's FC layer connected to the first Conv layer.
Thus, even if the number of total layers considered (one global layer + classifier) is smaller compared to the latter stages (multiple global layers + classifier), the model size (\ie~number of parameters) can be larger.

Indeed, we are aware that by training any of these models on CIFAR10~\cite{mcmahan2017communication} for more communication rounds, either the \oursystem or the regular end-to-end \fl can reach higher test accuracy such as 85\% with standard $FedAvg$.
However, the training rounds used here are sufficient for our needs, as our goal is to evaluate the performance of \oursystem (\ie~what is the cost for reaching the same accuracy), and not to achieve the best possible accuracy on this classification task.

\begin{table}[t!]
\small
\caption{Communication overhead (rounds and amount) of \oursystem to reach the same accuracy as end-to-end \fl system.}
\label{tab:overal_performance}
\begin{tabular}{lllll}
\hline
Model   & Data    & \begin{tabular}[c]{@{}l@{}}Baseline\\ Acc.$^{\alpha}$\end{tabular} & \begin{tabular}[c]{@{}l@{}}Comm.\\ Rounds\end{tabular} & \begin{tabular}[c]{@{}l@{}}Comm.\\ Amount\end{tabular} \\ \hline
LeNet   & IID     & 98.93\%             & 56 (0.37$\times$)$^\delta$    & 0.38 $\times$     \\ %\cline{2-6}
        & Non-IID & 97.06\%$^\epsilon$  & -                             & -                 \\\hline
AlexNet & IID     & 68.50\%             & 97 (0.65$\times$)             & 0.63 $\times$     \\ %\cline{2-6} 
        & Non-IID & 49.49\%             & 79 (0.53$\times$)             & 0.53 $\times$     \\ \hline
VGG9    & IID     & 63.09\%             & 171 (1.14$\times$)            & 2.87 $\times$     \\ %\cline{2-6} 
        & Non-IID & 46.70\%             & 36 (0.24$\times$)             & 0.60 $\times$     \\ \hline
\end{tabular}
\begin{flushleft}
\footnotesize{
$^\alpha$Acc.: Test accuracy of 150 communication rounds in end-to-end FL;\\ $^{\delta}1\times$ refers to no overhead;
$^\epsilon$\oursystem reaches a maximum of 95.99\%.} \hfill
\end{flushleft}
\end{table}

\begin{table}[t!]
\caption{Time duration of \fl phases in \emph{one communication round}, when training LeNet, AlexNet and VGG9 models with \oursystem and end-to-end (E2E) \fl.}
\label{tab:fl_duration}
\small
\setlength\tabcolsep{4pt}
\begin{tabular}{lrlllll}
\hline
\multicolumn{1}{c}{\multirow{2}{*}{Model}} & \multicolumn{1}{c}{\multirow{2}{*}{Method}} & \multicolumn{5}{c}{Duration of FL phases (s)} \\ \cline{3-7} 
\multicolumn{1}{c}{}                       & \multicolumn{1}{c}{}                        & B.cast$^\alpha$  & Training  & Upload & Aggr.$^\delta$ & Total  \\ \hline
\multirow{2}{*}{\STAB{\rotatebox[origin=c]{90}{LeNet\ \ }}}                                      & \multicolumn{1}{l}{E2E}                 & 4.520   & 2691.0    & 6.645  & 0.064 & 2702.3 \\ \cline{2-7} 
                                           & \multicolumn{1}{l}{PPFL}                    & 18.96   & 6466.2    & 7.535  & 1.887 & 6496.5 \\ \cline{3-7} 
                                           & \textit{- layer 1}                          & 4.117   & 1063.3    & 1.488  & 0.426 & 1069.8 \\
                                           & \textit{- layer 2}                          & 4.670   & 2130.6    & 1.627  & 0.692 & 2138.3 \\
                                           & \textit{- layer 3}                          & 5.332   & 2315.2    & 1.745  & 0.676 & 2323.6 \\
                                           & \textit{- clf.$^\epsilon$}                             & 4.845   & 957.16    & 2.675  & 0.093 & 964.87  \\ \hline
\multirow{2}{*}{\STAB{\rotatebox[origin=c]{90}{AlexNet\ \ }}}                                    & \multicolumn{1}{l}{E2E}                 & 14.58   & 3772.0    & 6.122  & 0.061 & 3792.8 \\ \cline{2-7} 
                                           & \multicolumn{1}{l}{PPFL}                    & 57.24   & 14236     & 16.89  & 3.290 & 14316  \\ \cline{3-7} 
                                           & \textit{- layer 1}                          & 16.20   & 2301.8    & 4.690  & 0.129 & 2322.9 \\
                                           & \textit{- layer 2}                          & 12.56   & 4041.1    & 4.777  & 0.174 & 4058.8 \\
                                           & \textit{- layer 3}                          & 10.31   & 4609.4    & 5.388  & 0.243 & 4625.6 \\
                                           & \textit{- clf.}                             & 18.17   & 3283.8    & 2.033  & 2.744 & 3309.5 \\ \hline
\multirow{2}{*}{\STAB{\rotatebox[origin=c]{90}{VGG9\ \ }}}                                       & \multicolumn{1}{l}{E2E}                 & 14.10   & 2867.1    & 8.883  & 0.067 & 2890.2 \\ \cline{2-7} 
                                           & \multicolumn{1}{l}{PPFL}                    & 353.5   & 21389     & 173.8  & 4.066 & 21924  \\ \cline{3-7} 
                                           & \textit{- layer 1}                          & 127.5   & 4245.7    & 95.58  & 0.375 & 4469.5 \\
                                           & \textit{- layer 2}                          & 77.22   & 2900.6    & 24.82  & 0.207 & 3003.1 \\
                                           & \textit{- layer 3}                          & 79.18   & 3703.1    & 24.84  & 0.223 & 3807.6 \\
                                           & \textit{- layer 4}                          & 27.05   & 2987.9    & 12.15  & 0.235 & 3027.6 \\
                                           & \textit{- layer 5}                          & 21.47   & 2404.4    & 9.137  & 0.347 & 2435.7 \\
                                           & \textit{- layer 6}                          & 10.95   & 2671.0    & 4.768  & 0.571 & 2687.9 \\
                                           & \textit{- clf.}                             & 10.11   & 2476.4    & 2.478  & 2.108 & 2493.2 \\ \hline
\end{tabular}
\begin{flushleft}
\footnotesize{
$^{\alpha}$B.cast: Broadcast;
$^\delta$Aggr.: Aggregation;
$^\epsilon$clf.: Classifier.} \hfill
\end{flushleft}
\end{table}

\mysubsubsection{Communication Duration of \fl Phases.}
In the next experiment, we investigate the wall-clock time needed for running \oursystem's phases in one communication round: broadcast of the layer from server to clients, training of the layer at the client device, upload the layer to the server, aggregate all updates from clients and apply $FedAvg$.
Depending on each layer's size and TEE memory size, batch size can start from 1 and go as high as the TEE allows.
However, since our models are uneven in layer sizes (with VGG9 being the largest), we set the batch size to 1 to allow comparison, and also capture an upper bound on the possible duration of each phase in each model training.
Indeed, we confirmed that increasing batch size for small models that allow it (\eg~AlexNet with batch size=16), incrementally reduces the duration of phases.

Table~\ref{tab:fl_duration} shows the break-down of time taken for each phase, for three models and two datasets (LeNet on MNIST; AlexNet and VGG9 on CIFAR10) and IID data.
As expected, layer-wise \fl increases the total time compared to end-to-end \fl because each layer is trained separately, but the previously trained and finalized layers still need to be processed in the forward pass.
In fact, these results are in line with the complexity analysis shown earlier in Sec.~\ref{sec:layer-wise-training-in-tees}, \ie~to finish the training of all layers, layer-wise training introduces a $3\times$ or higher delay, similar to the number of layers.
On the one hand, we argue that applications can tolerate this additional delay if they are to be protected from privacy-related attacks, despite the execution time increase being non-negligible and up to a few hours of training.
Indeed, models can be (re)trained on longer timescales (\eg~weekly, monthly), and rounds can have a duration of 10s of minutes, while being executed in an asynchronous manner.
On the other hand, training one layer in \oursystem costs similar time to the end-to-end \fl training of the complete model.
This highlights that the minimum client contribution time is the same as end-to-end \fl: clients can choose to participate in portions of an \fl round, and in just a few \fl rounds.
For example, a client may contribute to the model training for only a few layers in any given \fl round.

Among all \fl phases, local training costs the most, while the time spent in server aggregation and averaging is trivial, regardless if it is non-secure (\ie~end-to-end FL) or secure (\oursystem).
Regarding VGG9, layer-wise training of early layers significantly increases the communication time in broadcast and upload, because the Conv layers are with a small number of filters and consequently the following classifier's FC layer has a large size. 
This finding hints that selecting suitable \dnns to be trained in \oursystem (\eg~AlexNet vs. VGG9) is crucial for practical performance.
Moreover, and according to the earlier \fl performance results (also see Table~\ref{fig:greedy_acc}), it may not be necessary to train all layers to reach the desired ML utility.

%%%%%%%%%%%%%%%%%%%%%%%%%%%%%%%%%%%%
%%%%%%%%%%%% subsection %%%%%%%%%%%%
\subsection{Is the \oursystem ML Performance Comparable to State-of-art \fl?}
\label{sec:mlperformance-results}

\begin{figure}[t!]
    \centering
    \begin{subfigure}{1\columnwidth}
        \includegraphics[width=1\columnwidth]{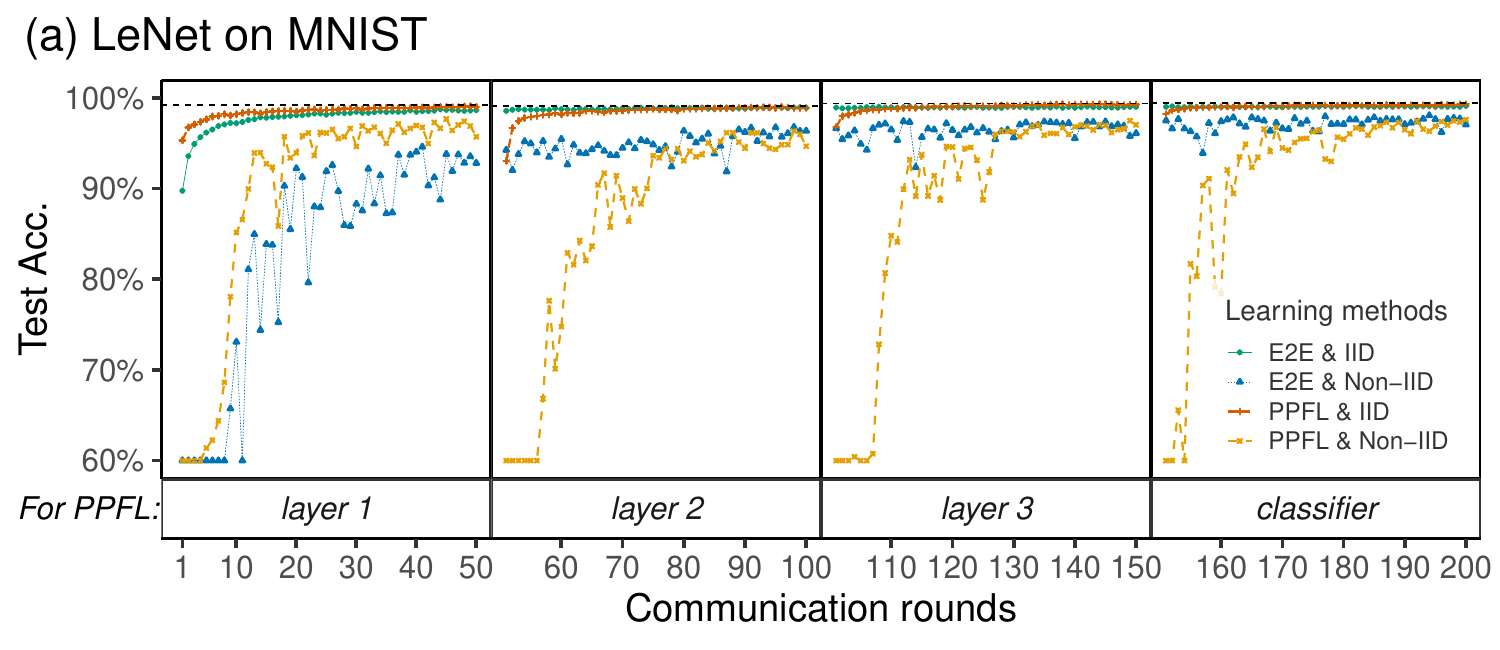}\vspace{-15pt}
        \label{fig:ppfl_acc_lenet}
    \end{subfigure}
    \begin{subfigure}{1\columnwidth}
        \includegraphics[width=1\columnwidth]{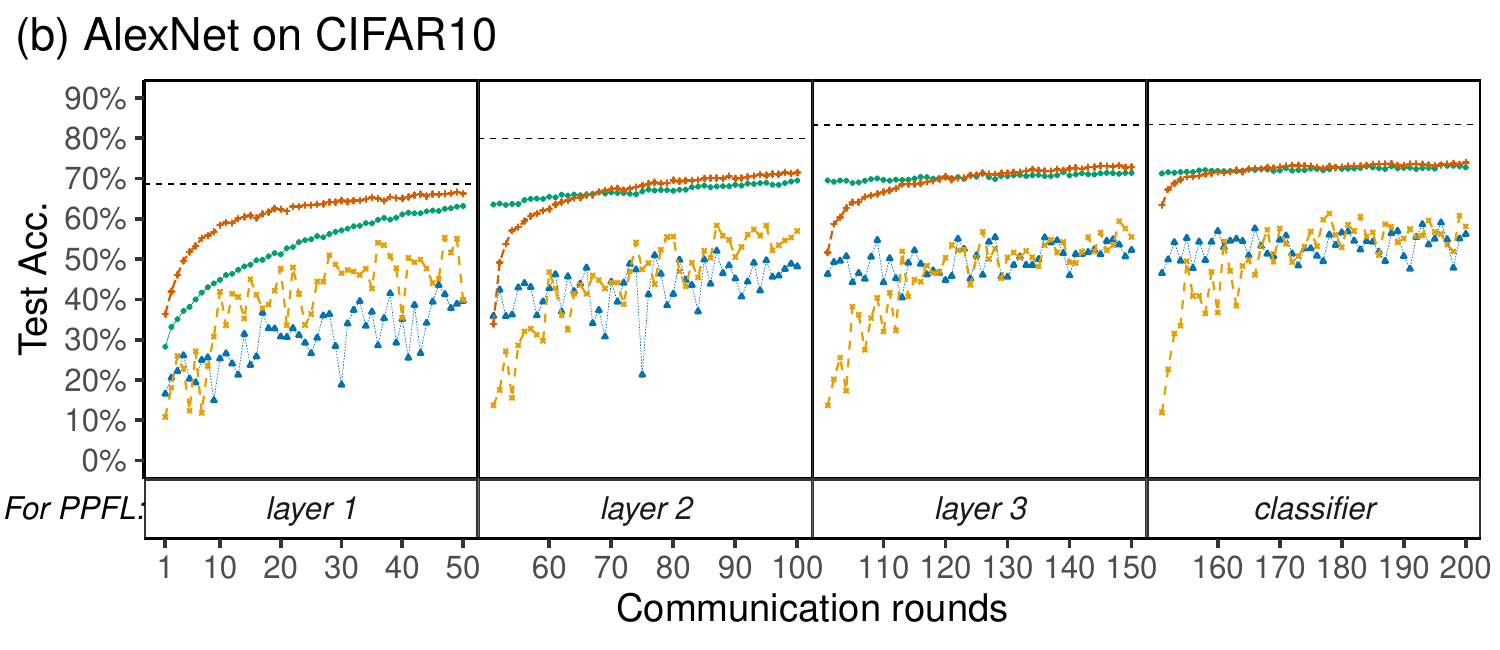}\vspace{-15pt}
        \label{fig:ppfl_acc_alexnet}
    \end{subfigure}
    \begin{subfigure}{1\columnwidth}
        \includegraphics[width=1\columnwidth]{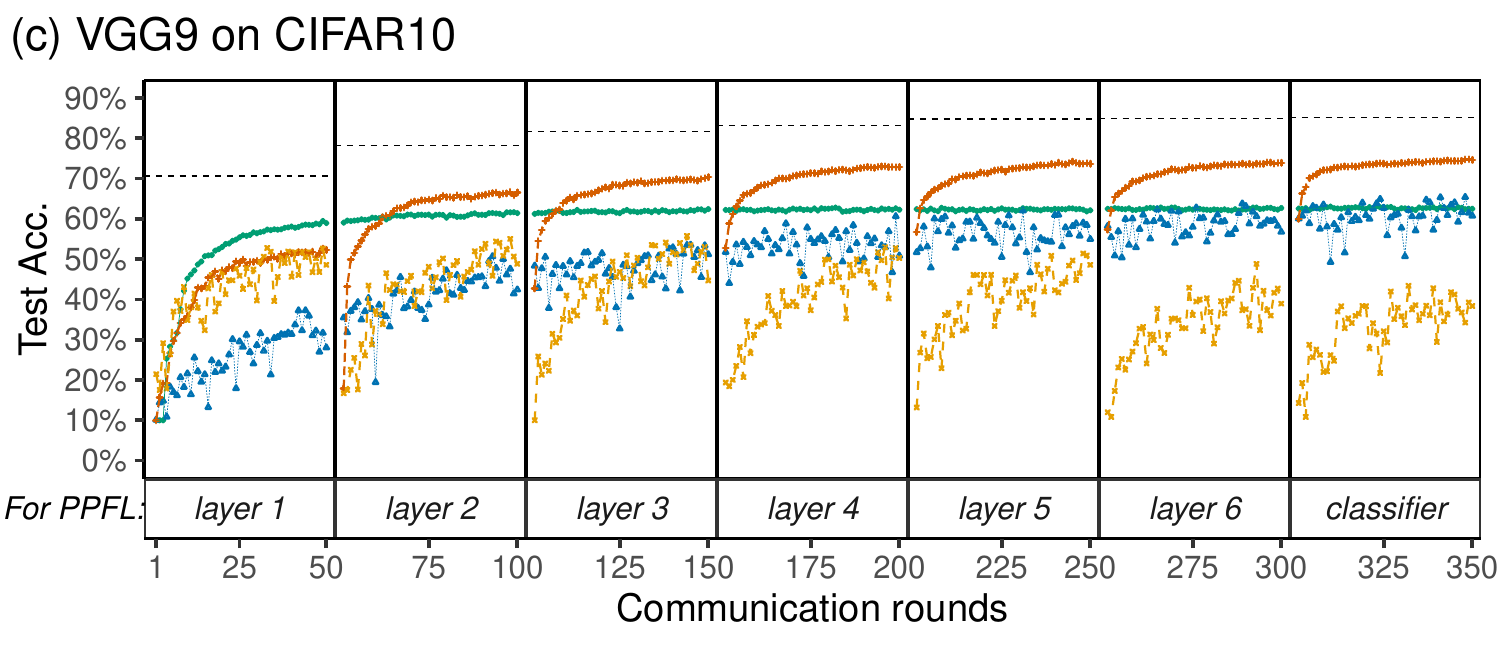}\vspace{-15pt}
        \label{fig:ppfl_acc_vgg}
    \end{subfigure}
    \caption{Test accuracy of training LeNet, AlexNet, and VGG9 models on IID and Non-IID datasets when using \oursystem. Horizontal dashed lines refer to the accuracy that the centralized training reaches after every 50 epochs.
    Note: end-to-end (E2E) \fl trains the complete model rather than each layer, and the `Layer No.' at x-axis are \emph{only} applicable to \oursystem.}
    \label{fig:greedy_acc}
\end{figure}

\begin{figure*}[t!]
    \centering
    \includegraphics[width=2\columnwidth]{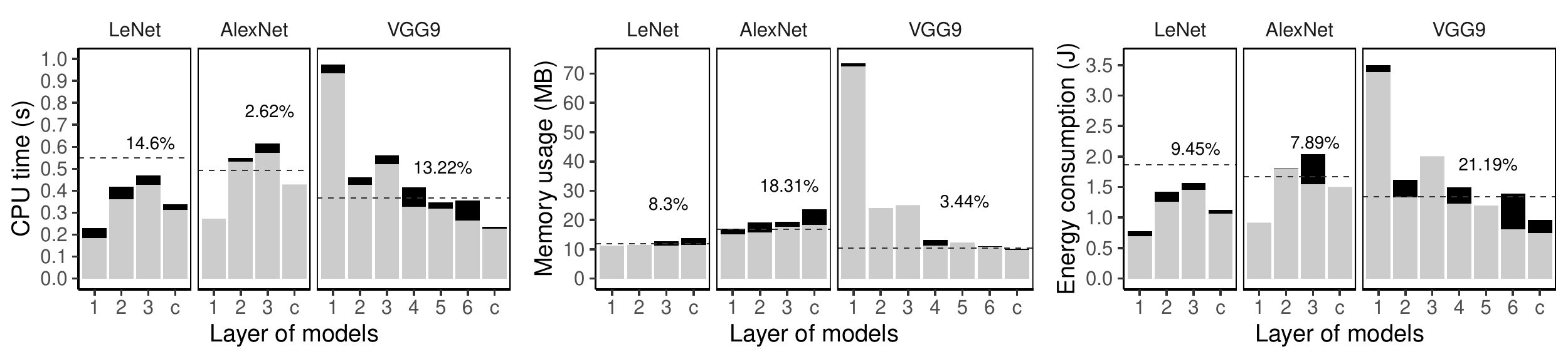}
    \caption{System performance of the client devices when training LeNet, AlexNet, and VGG9 using \oursystem, measured on \emph{one step of training} (\ie~one batch of data).
    The light grey bar (\lightgreybar) refers to learning without TEEs, and the black bar (\blackbar) refers to overhead when the layer under training is inside the TEE. Percentage (\%) of the overhead (averaged on one model) is shown above these bars. Horizontal dashed lines signify the cost of end-to-end \fl.
    In x-axis, `c' refers to `classifier'.}
    \label{fig:greedy_tee}
\end{figure*}

In these experiments, we reduce the number of communication rounds that each layer in \oursystem is trained to 50, finish the training process per layer, and compare its performance with centralized layer-wise training, as well as regular end-to-end \fl.
The latter trains the full model for all rounds up to that point.
For example, if \oursystem trains the first layer for 50 rounds, and then the second layer for 50 rounds, the end-to-end \fl will train all the model (end-to-end) for 100 rounds.

As shown in Figure~\ref{fig:greedy_acc}, training LeNet on the ``easy'' task of MNIST data (IID or not) leads quickly to high ML performance, regardless of the \fl system used.
Training AlexNet on IID and Non-IID CIFAR10 data can lead to test accuracy of 74\% and 60.78\%, respectively, while centralized training reaches 83.34\%.
Training VGG9, which is a more complex model on IID and Non-IID CIFAR10 data leads to lower performances of 74.60\% and 38.35\%, respectively, while centralized training reaches 85.09\%.
We note the drop of performance in \oursystem when every new layer is considered into training. 
This is to be expected, since \oursystem starts from scratch with the new layer, leading to a significant performance drop in the first \fl rounds.
Of course, towards the end of the 50 rounds, \oursystem performance matches and in some cases surpasses that of end-to-end \fl.

In general, with more layers being included in the training, the test accuracy increases.
Interestingly, in more complex models (\eg~VGG9) with Non-IID data, \oursystem can lead to a drop in ML performance when the number of layers keeps increasing.
In fact, in these experiments, it only reaches $\sim$55\% after finishing the second layer and drops.
One possible reason for this degradation is that the first layers of VGG9 are small and maybe not capable of capturing heterogeneous features among Non-IID data, which consequently has a negative influence on the training of latter layers.
On the other hand, this reminds us that we can have early exits for greedy layer-wise \oursystem on Non-IID data.
For example, clients that do not have enough data, or already have high test accuracy after training the first layers can quit before participating in further communication rounds.
Overall, the layer-wise training outperforms end-to-end \fl during the training of the first or second layer.

We further discuss possible reasons for \oursystem's better ML performance compared to end-to-end \fl.
On the one hand, this could be due to some DNN architectures (\eg~VGG9) being more suitable for layer-wise \fl.
For example, training each layer separately may allow \oursystem to overcome possible local optima at which the backward propagation can ``get stuck'' in end-to-end \fl.
On the other hand, hyper-parameter tuning may help improve performance in both layer-wise and end-to-end \fl, always with the risk of overfitting the data.
Indeed, achieving the best ML performance possible was not our focus, and more in-depth studying is needed in the future, to understand under what setups layer-wise can perform better than end-to-end \fl.

%%%%%%%%%%%%%%%%%%%%%%%%%%%%%%%%%%%%
%%%%%%%%%%%% subsection %%%%%%%%%%%%
\subsection{What is the \oursystem Client-Side System Cost?}
\label{sec:clientside-system-cost-results}

We further investigate the system performance and costs on the client devices with respect to CPU execution time, memory usage, and energy consumption.
Figure~\ref{fig:greedy_tee} shows the results for all three metrics, when training LeNet on MNIST, AlexNet and VGG9 on CIFAR10, on IID data.
The metrics are computed for one step of training (\ie~one batch of data).
More training steps require analogously more CPU time and energy, but do not influence memory usage since the memory allocated for the model is reused for all subsequent steps. 
Here, we compare \oursystem with layer-wise training without TEEs, to measure the overhead of using the TEE.
Among the trained models, the maximum overhead is $14.6\%$ for CPU time, $18.31\%$ for memory usage, and $21.19\%$ for energy consumption.
In addition, when training each layer, \oursystem has comparable results with end-to-end training (\ie~horizontal dashed lines in Figure~\ref{fig:greedy_tee}).

%%%%%%%%%%%%%%%%%%%%%%%%%%%%%%%%%%%%
%%%%%%%%%%%% subsection %%%%%%%%%%%%
\subsection{What is the \oursystem ML and System Costs if Blocks of Layers were Trained in Clients?}
\label{sec:blocks-results}

\begin{figure}[t!]
    \centering
    \begin{subfigure}{1\columnwidth}
        \includegraphics[width=1\columnwidth]{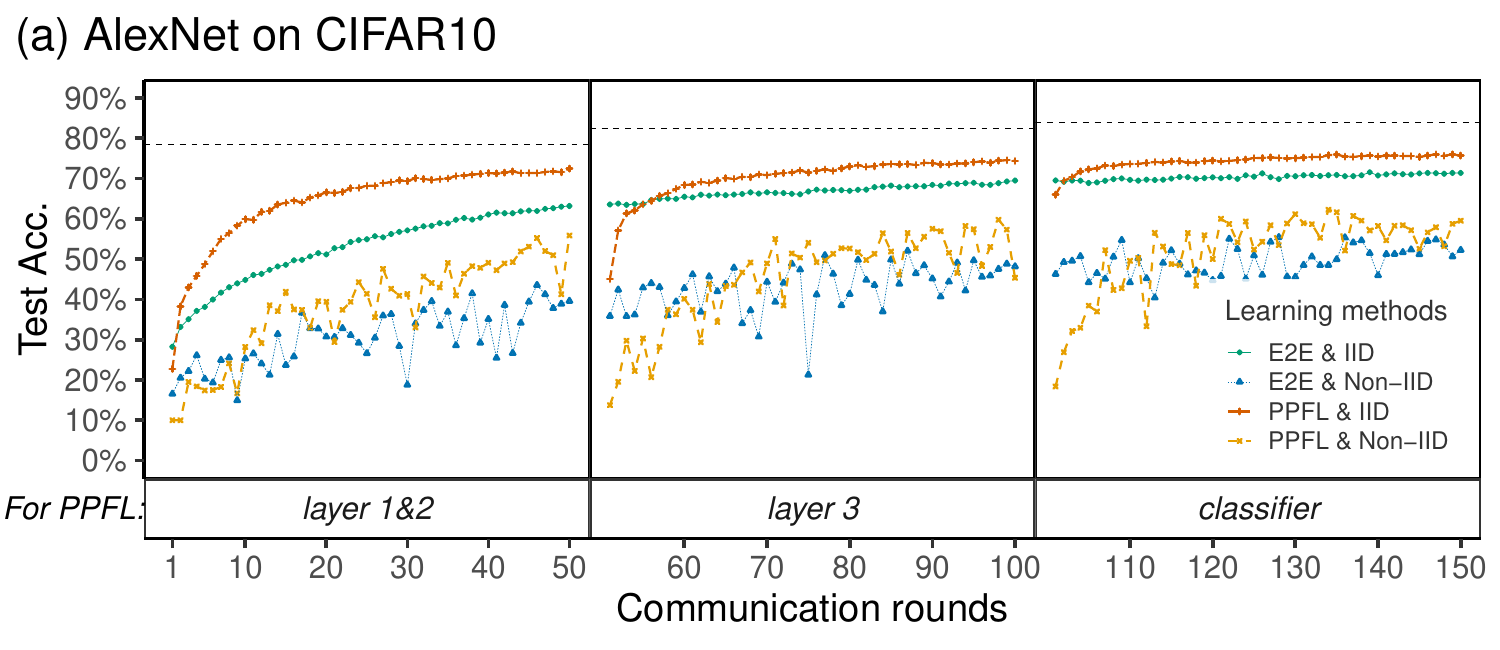}\vspace{-15pt}
        \label{fig:ppfl_acc_alexnet_b2}
    \end{subfigure}
    \begin{subfigure}{1\columnwidth}
        \includegraphics[width=1\columnwidth]{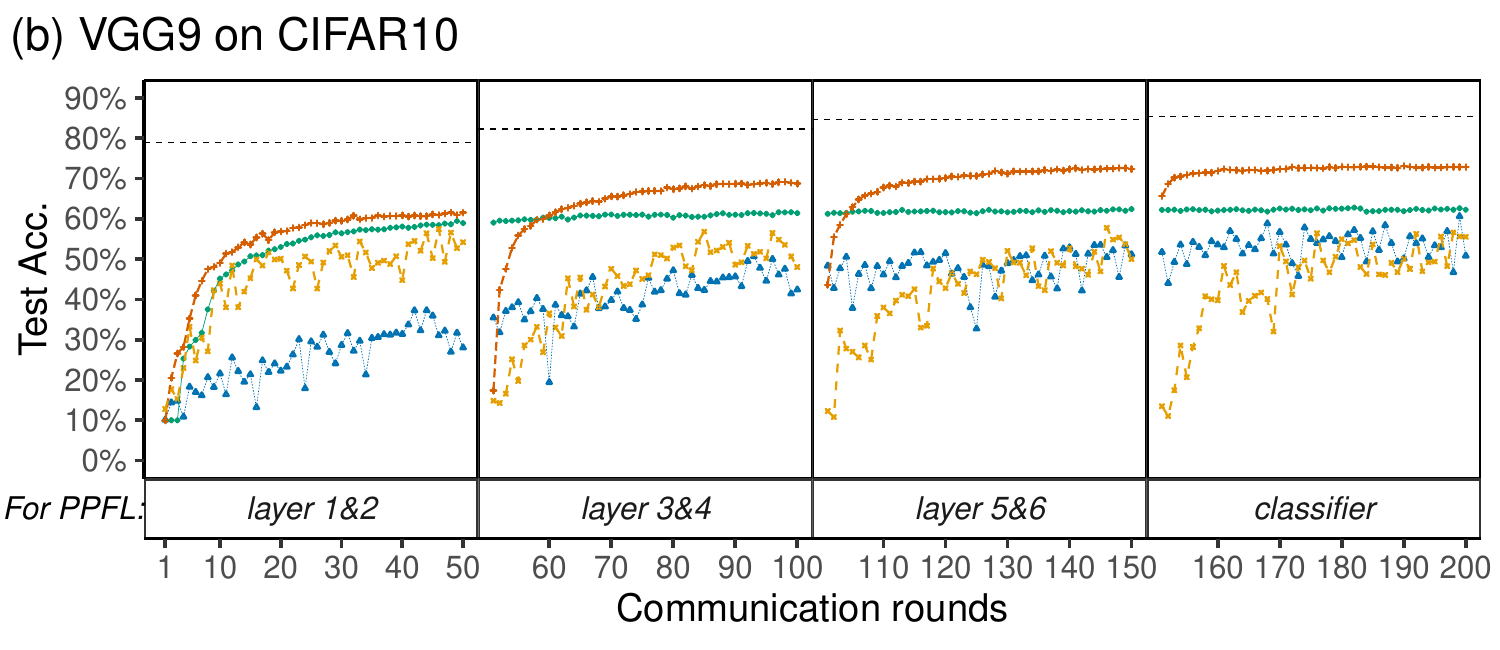}\vspace{-15pt}
        \label{fig:ppfl_acc_vgg_b2}
    \end{subfigure}
    \caption{Test accuracy of training AlexNet and VGG9 models on CIFAR10 (IID and Non-IID) when using \oursystem with blocks of two layers in TEE (Note: horizontal dashed lines refer to the accuracy that the end-to-end (E2E) \fl reaches after 50 communication rounds).}
    \label{fig:greedy_acc_b2}
\end{figure}

\begin{figure}[t!]
    \centering
    \includegraphics[width=1\columnwidth]{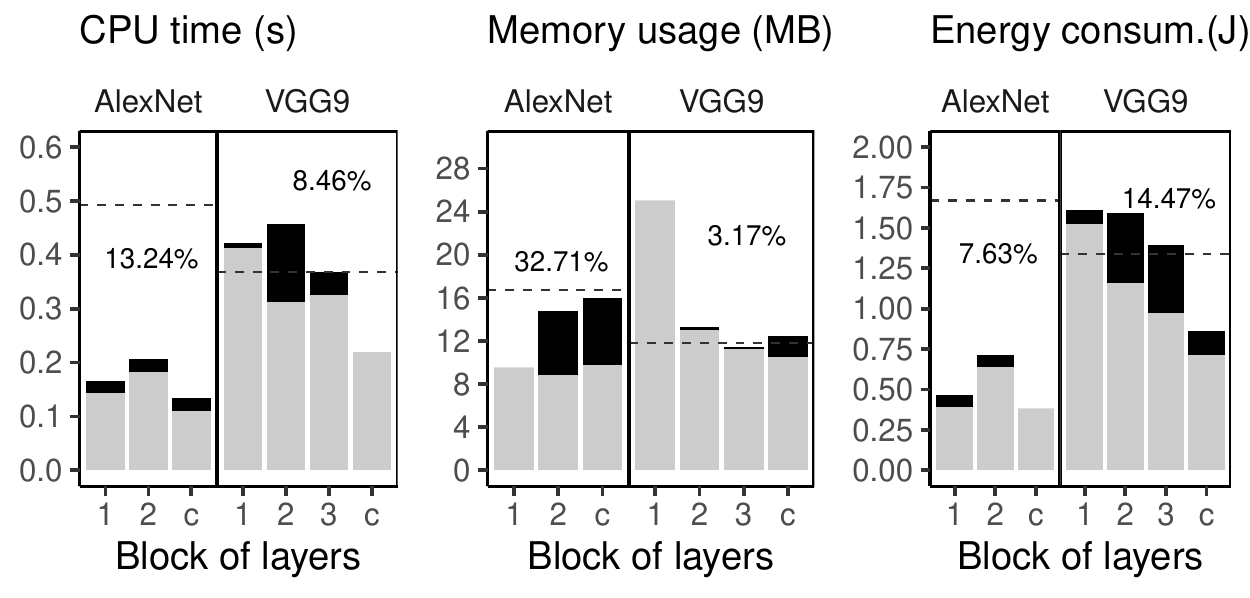}
    \caption{System performance of the client devices when training AlexNet and VGG9 models on CIFAR10 when using \oursystem with blocks of two layers in TEE (same settings as in Figure~\ref{fig:greedy_acc_b2}), measured on one step of training.
    The light grey bar (\lightgreybar) refers to learning without TEEs, and the black bar (\blackbar) refers to overhead when the block's layers under training are inside the TEE.
    Percentage (\%) of the overhead is shown above these bars.
    Horizontal dashed lines refer to the cost of end-to-end \fl.
    `c' refers to `classifier'.}
    \label{fig:greedy_tee_b2}
\end{figure}

As explained in Algorithm~\ref{algo:ppfl_server} of Sec.~\ref{sec:layer-wise-training-in-tees}, if the TEEs can hold more than one layers, it is also possible to put a block of layers inside the TEE for training.
Indeed, heterogeneous devices and TEEs can have different memory sizes, thus supporting a wide range of block sizes.
For these experiments, we assume all devices have the same TEE size and construct 2-layer blocks, and measure the system's test accuracy and ML performance on CIFAR10.
The performance of three or more layers inside TEEs could be measured in a similar fashion (if the TEE's memory can fit them).
We do not test LeNet on MNIST because it can easily reach high accuracy (around 99\%) as shown earlier and in previous studies~\cite{mcmahan2017communication, wang2020federated}.

Results in Figure~\ref{fig:greedy_acc_b2} indicate that training blocks of layers can reach similar or even better ML performance compared to training each layer separately (\ie~see Fig.~\ref{fig:greedy_acc}).
It can also improve the test accuracy of complex models such as VGG9, for which we noted a degradation of ML performance caused by the first layer's small size and incapacity to model the data (see Fig.~\ref{fig:greedy_acc}).
In addition, compared to training one layer at a time, training 2-layer blocks reduces the total required communication to reach the desired ML performance.
In fact, while aiming to reach same baseline accuracy as in Table~\ref{tab:overal_performance}, training 2-layer blocks requires half or less of communication cost than 1-layer blocks see Table~\ref{tab:overal_performance_2}.
Also, layer-wise training outperforms end-to-end \fl for similar reasons as outlined for Figure~\ref{fig:greedy_acc}.

\begin{table}[t!]
\small
\caption{Reduction of communication rounds and amount when training 2-layer instead of 1-layer blocks.}
\label{tab:overal_performance_2}
\begin{tabular}{lllcc}
\hline
Model   & Data    &  Comm. Rounds & Comm. Amount \\ \hline
AlexNet & IID  & $0.65\times \xrightarrow{} 0.18\times$   & $0.63\times \xrightarrow{} 0.27\times$    \\ 
        & Non-IID     & $0.53\times \xrightarrow{} 0.29\times$             & $0.53\times \xrightarrow{} 0.44\times$     \\ \hline
VGG9    & IID        & $1.14\times \xrightarrow{} 0.43\times$            & $2.87 \times\xrightarrow{} 1.07\times$     \\ 
        & Non-IID     &  $0.24 \times \xrightarrow{} 0.11\times$    & $0.60 \times \xrightarrow{} 0.27\times$      \\ \hline
\end{tabular}
\end{table}

Regarding the system cost, results across models show that the maximum overhead is 13.24\% in CPU time, 32.71\% in memory usage, and 14.47\% in energy consumption (see Fig.~\ref{fig:greedy_tee_b2}).
Compared to training one layer at a time, training layer blocks does not always increase the overhead.
For example, overhead when running VGG9 drops from 13.22\% to 8.46\% in CPU, from 2.44\% to 3.17\% in memory usage, and from 21.19\% to 14.47\% in energy consumption.
One explanation is that combining layers into blocks amortizes the cost of ``expensive'' with ``cheap'' layers.
Interestingly, \oursystem still has a comparable cost with end-to-end \fl training.

%%%%%%%%%%%%%%%%%%%%%%%%%%%%%%%%%%%%
%%%%%%%%%%%% subsection %%%%%%%%%%%%
\subsection{Can Bootstrapping the \oursystem with Public Knowledge Help?}
\label{sec:publicknowledge-results}

We investigate how the backend server of \oursystem can use existing, public models to bootstrap the training process for a given task.
For this purpose, we leverage two models (MobileNetv2 and VGG16) pre-trained on ImageNet to the classification task on CIFAR10. 
Because these pre-trained models contain sufficient knowledge relevant to the target task, training the last few layers is already adequate for a good ML performance.
Consequently, we can freeze all Conv layers and train the last FC layers within TEEs, thus protecting them as well.
By default, MobileNetv2 has one FC layer, and VGG16 has three FC layers at the end.
We test both cases that one and three FC layers are attached and re-trained for these two models, respectively.
CIFAR10 is resized to $224\times224$ in order to fit with the input size of these pre-trained models.
We start with a smaller learning rate of 0.001 to avoid divergence and a momentum of 0.9 because the feature extractors are well-trained.

\begin{figure}[t!]
    \centering
    \begin{subfigure}{0.5\columnwidth}
        \includegraphics[width=1\columnwidth]{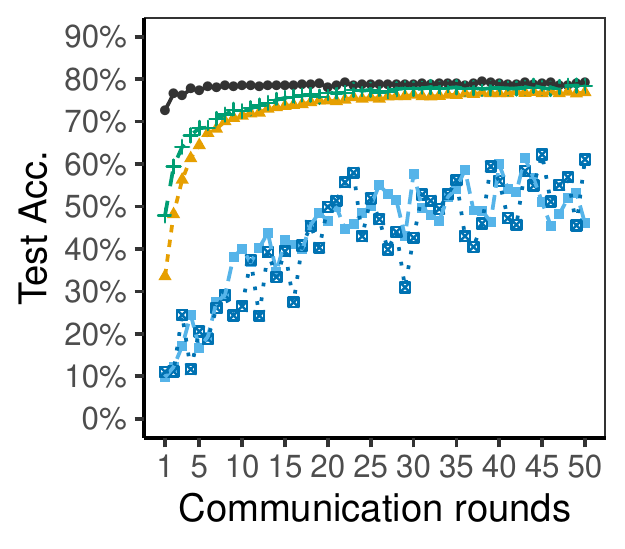}
        \caption{Transfer from MobileNetv2}
        \label{fig:ftl_acc_mobilenet}
    \end{subfigure}%
    \begin{subfigure}{0.5\columnwidth}
        \includegraphics[width=1\columnwidth]{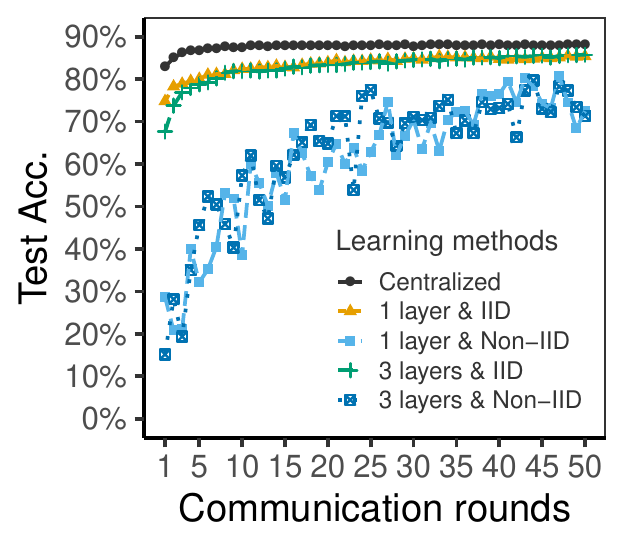}
        \caption{Transfer from VGG16}
        \label{fig:ftl_acc_vgg}
    \end{subfigure}
\caption{Test accuracy of training on CIFAR10 (IID and Non-IID) with public models MobileNetv2 and VGG16, pre-trained on ImageNet).
Both models are trained and tested with 1 and 3 FC layers attached at the end of each model.}
\label{fig:ftl_npmodel_acc}
\end{figure}

\mysubsubsection{Test Accuracy.}
Figure~\ref{fig:ftl_npmodel_acc} shows that the use of pre-trained first layers (\ie~feature extractors) to bootstrap the learning process can help the final \oursystem models reach test accuracy similar to centralized training.
Interestingly, transferring pre-trained layers from VGG16 can reach higher test accuracy than MobileNetv2.
This is expected because VGG16 contains many more \dnn parameters than MobileNetv2, which provides better feature extraction capabilities.
Surprisingly, attaching and training more FC layers at the end of any of the models does not improve test accuracy.
This can be due to the bottleneck of the transferred feature extractors, which since they are frozen, they do not allow the model to \emph{fully} capture the variability of the new data.

\mysubsubsection{Client-side System Cost.}
In order to measure client-side cost under this setting, we need to do some experimental adjustments.
The VGG16 (even the last FC layers) is too large to fit in TEEs.
Thus, we reduce the batch size to 1 and proportionally scale down all layers (\eg~from 4096 to 1024 neurons for one FC layer).
Indeed, scaling layers may lead to biases in results, but the actual performance cannot be worse than this estimation.
As shown in~\cite{mo2020darknetz}, larger models have less overhead because the last layers are relatively smaller compared to the complete size of the model.

\begin{figure}[t!]
    \centering
    \includegraphics[width=1\columnwidth]{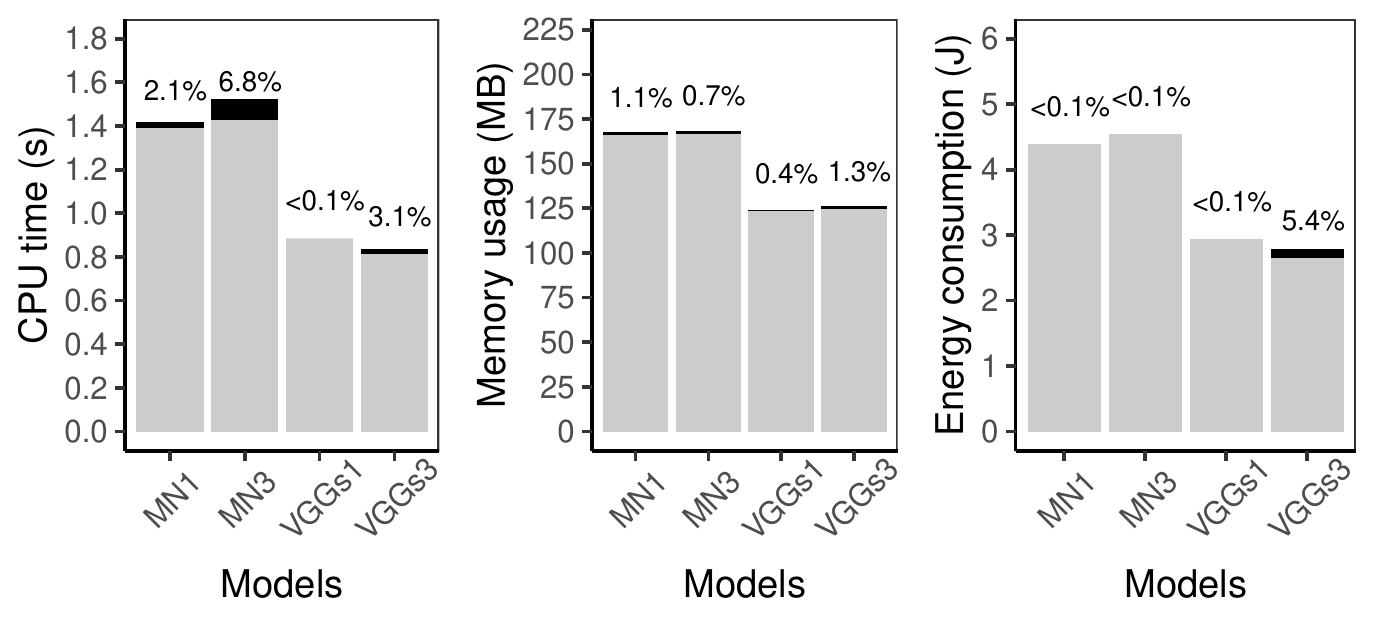}
    \caption{System performance of client devices when training with transferred public models on CIFAR10, measured on 1 step of training. Light grey bar (\lightgreybar): learning without TEEs; Black bar (\blackbar): overhead when layers under training are in TEE. Percentage (\%) of overhead shown above bars. MN1: MobileNetv2 with one layer for training (\ie~`1 layer' in Figure~\ref{fig:ftl_acc_mobilenet}). VGGs: a small size of VGG16.}
    \label{fig:ftl_npmodel_tee}
\end{figure}

Interestingly, results shown in Figure~\ref{fig:ftl_npmodel_tee} indicate that when we train and keep the last FC layers inside the client's on-device TEEs, there is only a small overhead incurred in terms of CPU time (6.9\%), memory usage (1.3\%), and energy consumption (5.5\%) in either model.
These results highlight that transferring knowledge can be a good alternative for bootstrapping \oursystem training and keep system overhead low.
In addition, we note that when the server does not have suitable public models, it is possible to first train a model on public datasets that have similar distribution with local datasets.
We refer to Appendix~\ref{sec:publicdataset-results} for more details on experimental results.

\section{Discussion \& Future Work}
\label{sec:discussion}

\mysubsubsection{Key Findings.}
\oursystem's experimental evaluation showed that:
\begin{itemize}[leftmargin=10pt]
    \item Protecting the training process (\ie~gradient updates) inside TEEs, and exposing layers only after convergence can thwart data reconstruction and property inference attacks. Also, keeping a model's last layer inside TEEs mitigates membership inference attacks.

    \item Greedy layer-wise \fl can achieve comparable ML utility with end-to-end \fl. While layer-wise \fl increases the total of communication rounds needed to finish all layers, it can reach the same test accuracy as end-to-end \fl with fewer rounds ($0.538\times$) and amount of communication ($1.002\times$).
    
    \item Most \oursystem system cost comes from clients' local training: up to $\sim$15\% CPU time, $\sim$18\% memory usage, and $\sim$21\% energy consumption in client cost when training different models and data, compared to training without TEEs.
    
    \item Training 2-layer blocks decreases communication cost by at least half, and slightly increases system overhead (\ie~CPU time, memory usage, energy consumption) in cases of small models.

    \item Bootstrapping \oursystem training process with pre-trained models can significantly increase ML utility, and reduce overall cost in communications and system overhead.
    
\end{itemize}

\mysubsubsection{Dishonest Attacks.}
The attacks tested here assume the classic `honest-but-curious' adversary~\cite{paverd2014modelling}.
      In \fl, however, there are also dishonest attacks such as backdoor~\cite{sun2019can, bagdasaryan2020backdoor} or poisoning attacks~\cite{fang2020local}, whose goal is to actively change the global model behavior, \eg~for surreptitious unauthorized access to the global model~\cite{jere2020taxonomy}.
In the future, we will investigate how TEEs' security properties can defend against such attacks.

\mysubsubsection{Privacy and Cost Trade-off.}
\oursystem guarantees `full' privacy by keeping layers inside TEEs.
However, executing computations in secure environments inevitably leads to system costs.
To reduce such costs, one can relax their privacy requirements, potentially increasing privacy risks due inference attacks with higher ``advantage''~\cite{zhao2020privacy-utility}.
For example, clients who do not care about high-level information leakages (\ie~learned model parameters), but want to protect the original local data, can choose to hide only the first layers of the model in TEEs.
We expect that by dropping clients already achieving good performance when training latter layers, we could gain better performance.
This may further benefit personalization and achieve better privacy, utility, and cost trade-offs.

\mysubsubsection{Model Architectures.}
The models tested in our layer-wise \fl are linear links cross consecutive layers. However, our framework can be easily extended to other model architectures that have been studied in standard layer-wise training.
For example, one can perform layer-wise training on (i) Graph Neural Networks by disentangling feature aggregation and feature transformation~\cite{you2020l2}, and (ii) Long Short-Term Memory networks (LSTMs), by adding hidden layers~\cite{sagheer2019unsupervised}.
There are other architectures that contain skipping connections to jump over some layers such as ResNet~\cite{he2016deep}.
No layer-wise training has been investigated for ResNets, but training a block of layers could be attempted by including the jumping shortcut inside a block.

\mysubsubsection{Accelerating Local Training.}
\oursystem uses only the CPU of client devices for local training. Training each layer does not introduce parallel processing on a device.
Indeed, more effective ways to perform this compute load can be devised. One way is that clients could use specialized processors (\ie~GPUs) to accelerate training.
\oursystem's design can integrate such advances mainly in two ways.
First, the client can outsource the first, well-trained, but non-sensitive layers, to specialized processors that can share computation and speed-up local training.
Second, recently proposed GPU-based TEEs can support intensive deep learning-like computation in high-end servers~\cite{hunt2020telekine, jang2019heterogeneous}.
Thus, such TEEs on client devices can greatly speed-up local training.
However, as GPU-TEE still requires small TCB to restrict attack surface, \oursystem's design can provide a way to leverage limited TEE space for privacy-preserving local training. 

\mysubsubsection{Federated Learning Paradigms.}
\oursystem was tested with $FedAvg$, but there are other state-of-art \fl paradigms that are compatible with \oursystem.
\oursystem leverages greedy layer-wise learning but does not modify the hyper-parameter determination and loss function (which have been improved in $FedProx$~\cite{li2018federated}) or aggregation (which is neuron matching-based in $FedMA$~\cite{wang2020federated}).
Compared with \oursystem that trains one layer until convergence, $FedMA$, which also uses layer-wise learning, trains each layer for one round, and then moves to the next layer.
After finishing all layers, it starts again from the first.
Thus, $FedMA$ is still vulnerable because gradients of one layer are accessible to adversaries.
\oursystem could leverage $FedMA$'s neuron-matching technique when dealing with heterogeneous (\ie~Non-IID) data~\cite{polifl}.
Besides, our framework is compatible with other privacy-preserving techniques (\eg~differential privacy) in FL.
This is useful during the model usage phase where some users may not have TEEs.
\oursystem can also be useful to systems such as FLaaS~\cite{kourtellis2020flaas} that enable third-party applications to build collaborative ML models on the device shared by said applications.
\section{Conclusion}
\label{sec:conclusion}

In this work, we proposed \oursystem, a practical, privacy-preserving federated learning framework, which protects clients' private information against known privacy-related attacks.
\oursystem adopts greedy layer-wise \fl training and updates layers always inside Trusted Execution Environments (TEEs) at both server and clients.
We implemented \oursystem with mobile-like TEE (\ie~TrustZone) and server-like TEE (\ie~Intel SGX) and empirically tested its performance.
For the first time, we showed the possibility of fully guaranteeing privacy and achieving comparable ML model utility with regular end-to-end \fl, without significant communication and system overhead.

\section{Acknowledgments}
We acknowledge the constructive feedback from the anonymous reviewers.
The research leading to these results received partial funding from the EU H2020 Research and Innovation programme under grant agreements No 830927 (Concordia), No 871793 (Accordion), No 871370 (Pimcity), and EPSRC Databox and DADA grants (EP/N028260/1, EP/R03351X/1).
These results reflect only the authors' view and the Commission and EPSRC are not responsible for any use that may be made of the information it contains.

%%%%%%%%%%%%%%%%%%%%%%%%%%%%%%%%%%%%%%%%%%%%
% Bibliography
\balance
\bibliographystyle{acm}
\bibliography{references}

%%%%%%%%%%%%%%%%%%%%%%%%%%%%%%%%%%%%%%%%%%%%
% appendix
\clearpage
\begin{appendix}
\section{Appendix}
\label{sec:appendix}

%%%%%%%%%%%%%%%%%%%%%%%%%%%%%%%%%%%%
%%%%%%%%%%%% subsection %%%%%%%%%%%%
\subsection{Transferring Public Datasets}
\label{sec:publicdataset-results}

The server can potentially gather data that have a similar distribution to clients' private data.
In initialization, the server trains a global model based on the gathered data rather than using one existing model.
Then, the server broadcasts the trained model to clients' devices.
Clients feed their private data into the model but update only the last layers inside the TEE during local training.
Also, only the last layers being trained are uploaded to the server for secure aggregation.
Because the server holds public data, we expect it to retrain the complete model before each communication round in order to keep fine-tuning the first layers.
Here, we fix the communication rounds to 20 and measure only the test accuracy. 
We expect the system cost to be similar to transferring from models because, similarly, only the last layers are trained at the client-side.

\begin{figure}[h!]
    \centering
    \begin{subfigure}{1\columnwidth}
        \includegraphics[width=1\columnwidth]{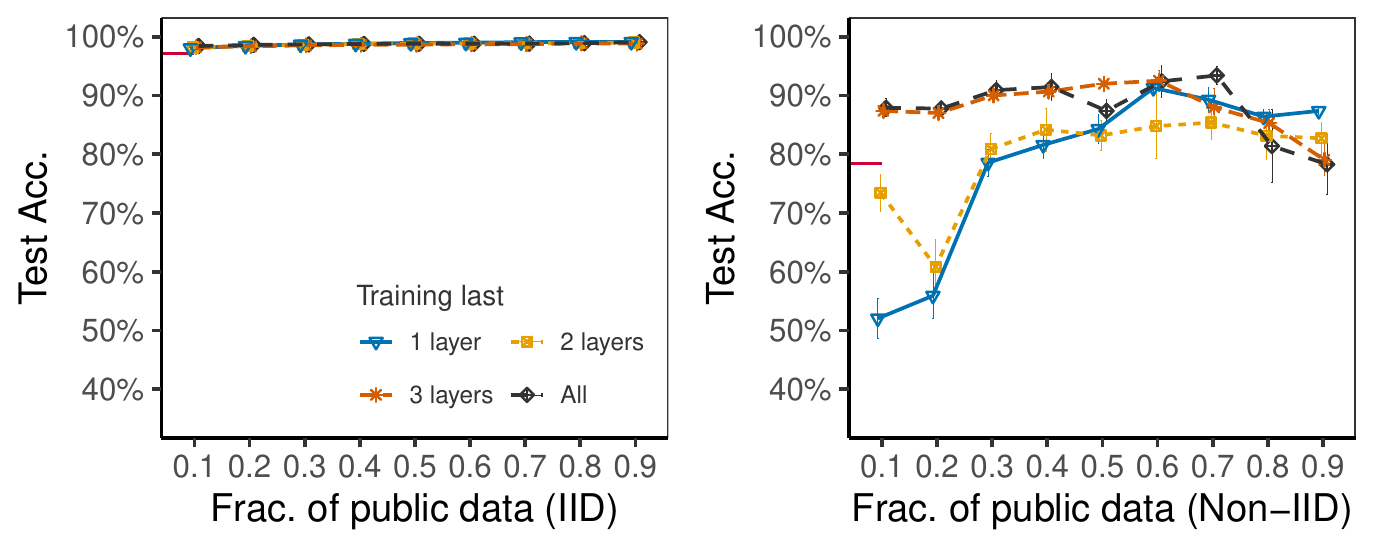}
        \caption{Transfer from public MNIST, to train LeNet.}
        \label{fig:ftl_acc_mnist}
    \end{subfigure}
    \begin{subfigure}{1\columnwidth}
        \includegraphics[width=1\columnwidth]{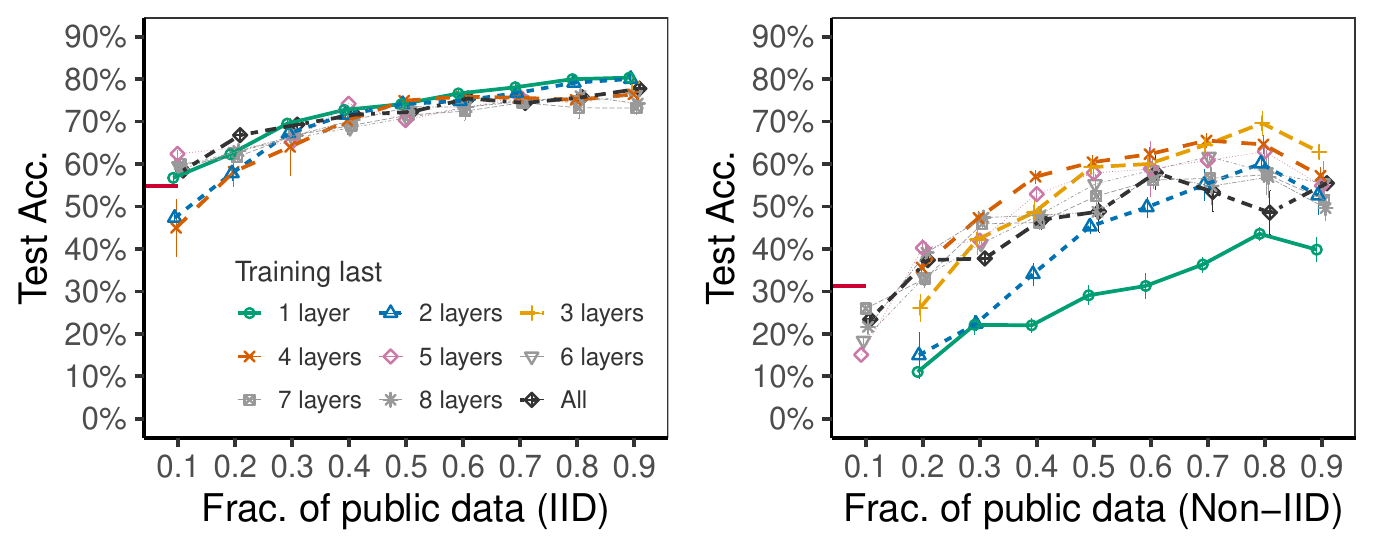}
        \caption{Transfer from public CIFAR10, to train VGG9.}
        \label{fig:ftl_acc_cifar}
    \end{subfigure}
\caption{Test accuracy when learning with public datasets. The short red line (\shortredline) starting from y-axis refers to end-to-end \fl. Each trail runs for 10 times, and error bars refer to 95\% confidence interval (Note: In the top left figure, test accuracy is very high and almost the same, as the range of y-axis is set as the same for the same dataset (\ie~MNIST here). In the bottom right figure (\ie~for CIFAR10), several trails fail to train and thus corresponding points are not plotted).}
\label{fig:ftl_npdata_acc}
\end{figure}

Test accuracy results are shown in Figure~\ref{fig:ftl_npdata_acc}.
It is indicated that in general when the server holds more public data, the final global model can reach a higher test accuracy.
This is as expected since the server gathers a larger part of the training datasets.
With complete training datasets, this process will finally become centralized training.
Nevertheless, this indication is not always held.
For example, in the IID case (see the two left plots in Figure~\ref{fig:ftl_npdata_acc}), when training all layers, servers with public data of 0.1 fraction outperform servers without public data, \ie~the end-to-end \fl, while regarding Non-IID of CIFAR10, servers with 0.1 fraction cannot outperform that without public data (see right plots in Figure~\ref{fig:ftl_acc_cifar}).
One reason for it is that the first layers, which are trained on public datasets, cannot represent all features of privacy datasets.
We also observe that when the server does not have enough public data (\eg~0.1 fraction), training only the last 1 or 2 layers can lead to extremely low performance or even failure.
Still, this is because the first layers cannot represent sufficiently the clients' datasets.

Another observation is that the number of training last layers does not have a significant influence on test accuracy in terms of IID cases, especially when the server holds more public data.
This is because learning from IID public data is able to represent the feature space of the complete (private) training datasets.
However, the results change when it comes to the Non-IID case.
The number of training last layers has a significant influence on test accuracy.
For instance, regarding VGG9, training only the last 1 or 2 layers at the client-side performs much worse compared to training 3 or 4 layers (see right plots in Figure~\ref{fig:ftl_acc_cifar}).
Moreover, training 3 or 4 layers tend to have better test accuracy than training more layers (\eg~all layers).
One explanation is that the feature extraction capability of first layers is good enough when the server has many public data, so fine-tuning these first layers at the client (\eg~training all layers) may destroy the model and consequently drop the accuracy.

Overall, by training only the last several layers at the client-side, \oursystem with public datasets can guarantee privacy, and in the meanwhile, achieve better performance than that of training all layers.
\end{appendix}

\end{document}